%% file: paper.tex
\documentclass[acmtog]{acmart} % official paper - also comment out vspace in abstract to avoid messing up the layout!
\setcitestyle{square}

\usepackage{booktabs}  
\usepackage{bm}  
\usepackage{enumitem}  
\usepackage[normalem]{ulem}

\usepackage[ruled]{algorithm2e} % For algorithms

\SetAlFnt{\small}
\SetAlCapFnt{\small}
\SetAlCapNameFnt{\small}
\SetAlCapHSkip{0pt}
\IncMargin{-\parindent}

\newcommand{\changed}[1]{\textcolor[rgb]{0,0,0}{#1}}

\newcommand{\systemName}{\textsc{FrankenGAN}\xspace}

\setcopyright{none} 

\begin{document}

% Title portion
\title{FrankenGAN: Guided Detail Synthesis for Building Mass Models Using Style-Synchonized GANs}

% Style-synchonized guided detail synthesis for Architectural Models
% Urban Greebler: Data-driven detail synthesis
% FrankenGAN: Directable Urban Texturing
% Data-driven Detailing of Urban Areas
% 
% 

%\author{papers\_325}
%\renewcommand{\shortauthors}{papers\_325}

\author{Tom Kelly}
\affiliation{\institution{University College London}}
\affiliation{\institution{University of Leeds}}
%\email{twakelly@gmail.com}
\authornote{joint first authors}

\author{Paul Guerrero}
\affiliation{\institution{University College London}}
\authornotemark[1]
%\email{paul.guerrero@ucl.ac.uk}

\author{Anthony Steed}
\affiliation{\institution{University College London}}
%\email{a.steed@ucl.ac.uk}

\author{Peter Wonka}%{$^\dagger$}
\affiliation{\institution{KAUST}}
\authornote{joint last authors}
%\email{peter.wonka@kaust.edu.sa}

\author{Niloy J. Mitra}
\affiliation{\institution{University College London}}
\authornotemark[2]
%\email{n.mitra@ucl.ac.uk}

%\authorsaddresses{email: twakelly@gmail.com; paul.guerrero@ucl.ac.uk; a.steed@ucl.ac.uk; peter.wonka@kaust.edu.sa; n.mitra@ucl.ac.uk}

%\authorsaddresses{Tom Kelly: twakelly@gmail.com; Paul Guerrero: paul.guerrero@ucl.ac.uk; Anthony Steed: a.steed@ucl.ac.uk; Peter Wonka: peter.wonka@kaust.edu.sa; Niloy J. Mitra: n.mitra@ucl.ac.uk}

\begin{abstract}
Coarse building mass models are now routinely generated at scales ranging from individual buildings to whole cities. Such models can be abstracted from raw measurements, generated procedurally, or created manually. However, these models typically lack any meaningful geometric or texture details, making them unsuitable for direct display. We introduce the problem of automatically and realistically decorating such models by adding semantically consistent geometric details and textures. Building on the recent success of generative adversarial networks~(GANs), we propose \systemName, a cascade of GANs that creates plausible details across multiple scales over large neighborhoods. The various GANs are synchronized to produce consistent style distributions over  buildings and neighborhoods. We provide the user with direct control over the variability of the output. We allow him/her to interactively specify the style via images and manipulate style-adapted sliders to control style variability. We test our system on several large-scale examples. The generated outputs are qualitatively evaluated via a set of perceptual studies and are found to be realistic, semantically plausible, and consistent in style.
%\vspace{40pt}
\end{abstract}




\begin{teaserfigure}
  \vspace{-5pt}
  \centering
  \includegraphics[width=\textwidth]{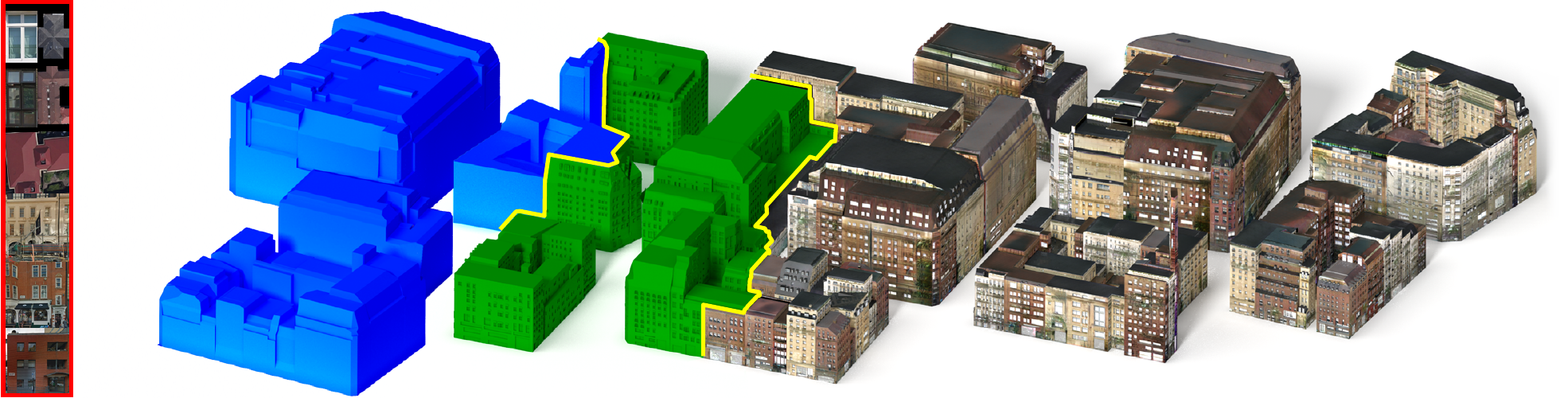}
  \caption{We introduce \systemName, a method to add detail to coarse mass models.
  %This method automatically adds diverse geometry and texture detail to
  This method allows the user to automatically generate diverse geometry and texture detail for
  %can automatically be added to
   a mass model (blue), while also giving the user control over the resulting style (through exemplar images, red). Geometric details (green) and textures (right) are generated by multiple generative adversarial networks with synchronized styles. A detailed view of this model is shown in Figure~\ref{fig:london}.}
  \label{fig:teaser}
\end{teaserfigure}

\maketitle

\input{sections/introduction.tex}

\input{sections/relatedWork.tex}
\input{sections/overview.tex}
\input{sections/method.tex}

\input{sections/results.tex}
\input{sections/conclusion.tex}

% Bibliography
\bibliographystyle{ACM-Reference-Format}
\bibliography{styleSynchronized,BibtexPeterWonkaMyPaper,gan}

%\appendix

%\input{sections/index.tex}

\end{document}

%% file: sections/introduction.tex
\section{Introduction}

%\the\abovecaptionskip

% Urban modeling has made tremendous progress in the last decades by building on the availability of large-scale raw data (e.g., aerial images, Lidar scans, etc.) and advances in supervised machine learning algorithms. As a result, abstracted \textit{mass models} over large neighborhoods can automatically be created by analyzing aerial images and/or Lidar scans~\cite{Kelly:SIGA:2017}. In addition to being light weight, such models are often semantically structured, geographically tagged, and provide plausible abstractions of urban facades. 

We propose a framework to add geometric and texture details to coarse building models, commonly referred to as  {\em mass models}. There are multiple sources of such coarse models: they can be generated from procedural models, reconstructed from aerial images and \mbox{LIDAR} data, or created manually by urban planners, artists and architects. %designing  hypothetical cityscapes. % proposed or fictional new development.

% Urban modeling has made tremendous progress in the last decades by building on the availability of large-scale raw data (e.g., aerial images, Lidar scans, etc.) and advances in supervised machine learning algorithms. As a result, abstracted \textit{mass models} over large neighborhoods can automatically be created by analyzing aerial images and/or Lidar scans~\cite{Kelly:SIGA:2017}. In addition to being light weight, such models are often semantically structured, geographically tagged, and provide plausible abstractions of urban facades. 

% \paul{I think we should not pitch it too much as a reconstruction paper, we should make it clear that mass models can come from various sources (including manual and procedural generation) and need not come from reconstruction. The argument in the next paragraph alone may otherwise not be very convincing, since attaching rectified images can give results that are nicer than ours. Since we will probably detect the detail labels from the final images, the details would be just as good (or maybe better) than ours. The downside is that this is only possible if we want to reconstruct a real building.}

\begin{figure}[t!]
    \centering
    \includegraphics[width=\columnwidth]{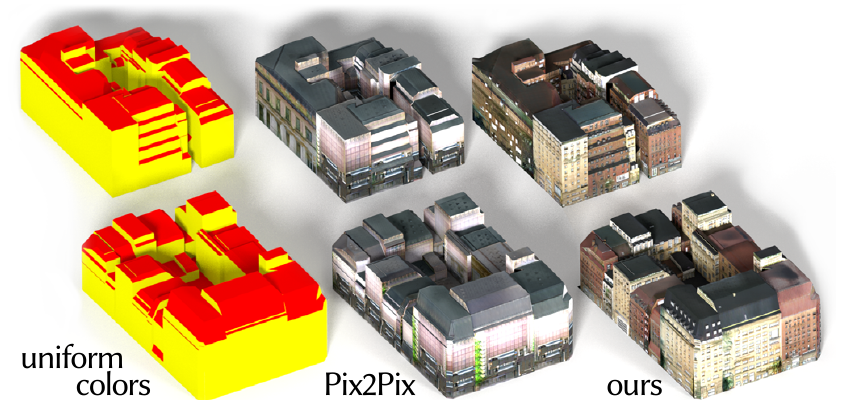}
    \vspace{-10pt}
    \caption{Naive options to texture a mass model give unconvincing results. Pix2Pix shows heavy mode collapse, discussed in Section~\ref{sec:qualitative_comparisons}.}
    \label{fig:intro_naive}
    \vspace{-5pt}
\end{figure}

By themselves, these mass models lack geometric and texture details. They therefore look unrealistic when directly displayed.
For many applications, decorating these models by adding details that are automatically generated would be desirable. However, naive attempts to decorate mass models lead to unconvincing results. Examples include assigning uniform colors to different building fronts, `attaching' rectified street-view images (when available) to the mass model faces, or synthesizing facade textures using current generative adversarial networks (see Figure~\ref{fig:intro_naive}). A viable alternative is to use  rule-based procedural models, for both geometric and texture details. However, such an approach requires time and expertise. 
Instead, we build on the recent success of machine learning using generative adversarial networks (GANs) to simplify the modeling process and to learn facade details directly from real-world data. We focus on two issues 
%There are two issues with current work that we would like to improve upon
to make GANs suitable for our application.
First, current GANs are concerned only with the generation of textures, but we also require
semantic and geometric details
for our urban modeling applications.
Second, current GANs provide little \textit{style control}. We would like to improve control over the style of facade details. For example, a user may wish to decorate a mass model in the style of a given example facade, or to specify how similar facades should be within an urban neighborhood. %automatically decorated block of mass models.
% \paul{possible additional challenge with directly using GANs: facades have a complex structure that is hard to generate end-to-end. (Current end-to-end results from blank facade to an image often look worse than ours since they don't have window layout ground truth as guidance.)}

%making the synthesis of details with realistic variation across different styles difficult.
%\twak{Manual texturing is time-consuming, even for skilled artists: UV coordinates must be defined, as well as specular, [normal], and texture bitmaps. Manual texturing, and procedural modeling are highly subjective, ignoring the large datasets available today.} Such an assisted workflow does not scale to detail large-scale models and across different styles. 
% (e.g. Sylvains paper\cite{xx})

\begin{figure*}[t!]
    \centering
    \includegraphics[width=\textwidth]{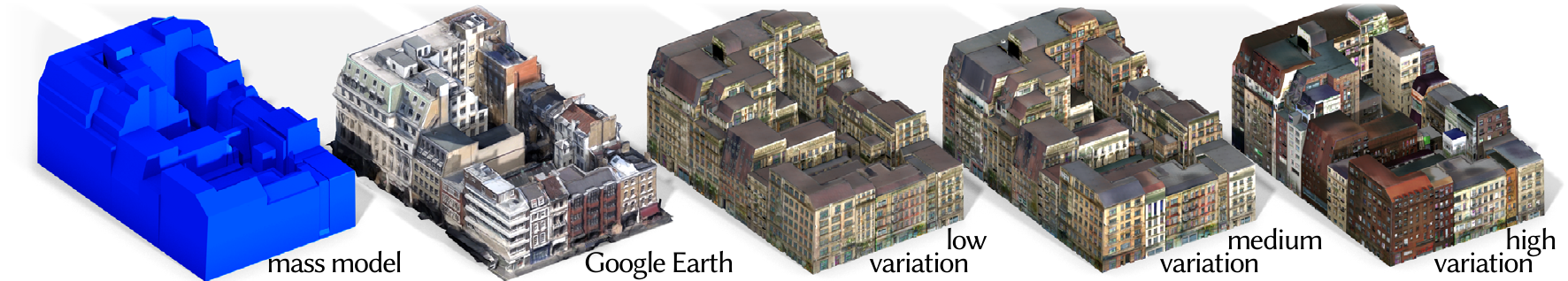}
    \caption{In contrast to photogrammetric reconstruction (second column), our method can be used to synthesize new facade layouts and textures. Style and variation can be controlled by the user; in columns 3 to 5, we show details generated by \systemName with low to high style variation.}
    \label{fig:photogrammetric}
\end{figure*}

Here, we consider the problem of automatically and realistically {\em detailing} mass models using user-supplied building images for style-guidance, with the option to adjust style variety. By details, we refer to both geometric and texture, details. Geometric details include balconies, window frames, roof types, chimneys, etc., while texture details refer to realistic facade appearances that are consistent with (latent) semantic elements such as windows, sills, etc. Further, we require the added details to be stylistically consistent and guided by user-provided input image(s). We do {\em not} expect the input images to be semantically-annotated or to correspond to the target mass models. 
The output of our algorithm (see Figure~\ref{fig:teaser}) is a mixture of 2.5D geometry with textures and realistic variations.

%\twak{ instead of being procedurally  instantiated. - balconies are procedural... }

Technically, we perform detailing in stages via a cascade of GANs. We engineer the individual GANs to generate particular styles that are encoded as latent vectors. As a result, we can synchronize the different GAN outputs by selecting style vectors from appropriate distributions. We demonstrate how to perform such style-guided synthesis for both geometric and texture details. By allowing style vectors to be guided by input images, we allow users to perform, at interactive rates, drag-and-drop stylization by simply providing  different target images  (see supplementary video).  
Our system, called \systemName, ensures that the resultant detailed models are realistic and are stylistically consistent both within individual buildings and between buildings in a neighborhood.

%para 5: demonstrate on various datasets .. 
We test our system by detailing a variety of mass models over large-scale city neighborhoods. We then compare the quality of the detailed models against baseline alternatives using a perceptual study. In summary, our main contributions are: 
(i)~introducing the problem of realistically {\em detailing} mass models with semantically consistent geometric and texture details; 
(ii)~presenting \systemName as an interactive system that utilizes latent style vectors via a cascade of synchronized GANs guided by examplar style images; and 
(iii)~demonstrating the system on several large-scale examples and qualitatively evaluating the generated output.
\changed{Source code and pre-trained networks are available at \emph{\url{http://geometry.cs.ucl.ac.uk/projects/2018/frankengan/}}.}

%% file: sections/relatedWork.tex
\section{Related Work}

In the following, we review related work on the computational design of facade layouts and generative adversarial networks.

\paragraph*{Computational facade layouts.}
Rule-based procedural modeling can be used to model facades and mass models~\cite{Wonka:2003:IA,Mueller:2006:PMB,Schwarz:2015:APM}. An alternative to pure procedural modeling is the combination of optimization with declarative or procedural descriptions. Multiple recent frameworks specifically target the modeling of facades and buildings~\cite{Bokeloh:2012:AMP,Lin:2011:SPR,Bao:2013:PFV,Ilcik:2015:LBP,Dang:2014:SAF}, and urban layouts~\cite{Vanegas:2012:IDU}. There are also multiple approaches that target more general procedural modeling~\cite{Talton:2011:MPM,Yeh:2013:STP,Ritchie:2015:CPM}.

\changed{Several methods obtain a facade layout by segmenting existing facade images, and then either fitting a procedural model~\cite{Teboul:PAMI:2013}, or optimizing the segmentation to follow architectural principles~\cite{Mathias:IJCV:2016,Cohen:CVPR:2014}. This reconstruction is different from our \emph{generative} approach, which synthesizes novel detail layouts without reference images.}

Another important avenue of recent work is the combination of machine learning and procedural modeling. One goal is inverse procedural modeling, where grammar rules and grammar parameters are learned from data. One example approach is Bayesian model merging~\cite{Stolcke:1994:IPG}, which was adopted by multiple authors for learning grammars~\cite{Talton:2012:LDP,Martinovic:2013:BGL}. While this approach shares some goals with our project, the learning techniques employed were not powerful enough to encode design knowledge and thus only very simple examples were presented.
Recently, deep learning was used for shape and scene synthesis of architectural content. Nishida et al.~\shortcite{Nishida:2016:ISU:2897824.2925951} proposed an interactive framework to interpret sketches as the outer shell of 3D building models. More recently, Wang et al.~\shortcite{Wang:2018:DCP} used deep learning to predict furniture placement inside a room. Automatic detailing of mass models, which is the focus of our method, has not yet been attempted using a data-driven approach. 

\paragraph*{Generative adverserial networks (GANs).}
Supervised deep learning has played a crucial role in recent developments in several computer vision tasks, e.g., object recognition \cite{he2016deep,krizhevsky2012imagenet}, semantic segmentation \cite{long2015fully}, and activity recognition/detection \cite{escorcia2016daps, tran2015learning}. 
In contrast, weakly supervised or unsupervised deep learning has been popular for image and texture synthesis tasks. 
In this context, Generative Adversarial Networks (GANs) have emerged as a promising family of unsupervised learning techniques that have recently been used to model simple natural images (e.g., faces and flowers) \cite{GAN}. They can learn to emulate the training set,  enable sampling from that domain and use the  learned knowledge for useful applications. Since their introduction, many variations of GANs have been proposed to overcome some of the impediments they face (e.g., instability during training)\\ \cite{WGAN,BEGAN,LaplacianPyrmid,Bigan,inception-GANs2,catGAN,improvGANfeature,zhao2016energy}. Three versions of GANs are of particular interest for our work. First, the Pix2Pix framework using conditional GANs~\cite{pix2pix} is useful for image to image translation. The authors provide results for translating a facade label image into a textured facade image. Second, CycleGAN~\cite{cycleGAN} is useful to learn image-to-image translation tasks \textit{without} requiring a corresponding pair of images in the two styles. Third, BicycleGAN~\cite{zhu2017multimodal} is another extension of image to image translation that improves the variations that can be generated.

Interestingly, GANs have proven useful in several core image processing and computer vision tasks, including image inpainting \cite{inpainting}, style transfer \cite{style-transfer}, super-resolution \cite{pix2pix,superresolution}, manifold traversing \cite{manifold-manipulation}, hand pose estimation \cite{hand-pose}, and face recognition \cite{face-recog}. However, the applications of GANs are not limited to the computer vision and image processing communities; adversarial models are being explored for graphics applications. Examples include street network synthesis \cite{StreetGAN}, volume rendering engines \cite{Berger:TVCG:2018}, and  adaptive city-scene generation \cite{sceneGAN}. In this paper, we adapt GANs to detail synthesis for detailing building mass models using exemplar images for style guidance.

% {\bf Paper wish list.}
% Procedural sketching SG 2016
% progressive GANs
% indoor modeling using deep learning - https://kwang-ether.github.io/pdf/deepsynth.pdf

%% file: sections/overview.tex
\begin{figure*}[t]
    \centering
    \includegraphics[width=\textwidth]{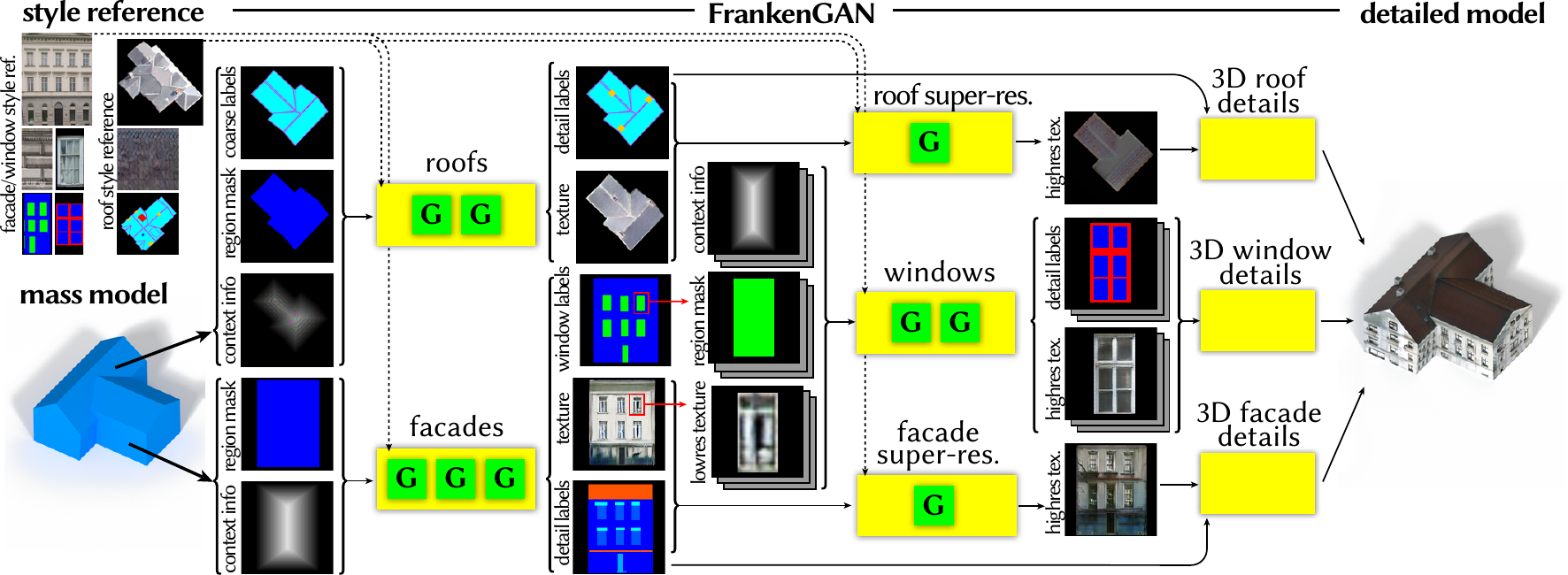}
    \caption{\systemName overview. Individual GANs are denoted by G, and yellow rectangles denote GAN chains or geometry generation modules. Given a mass model and an optional style reference (any part can be replaced by a random style), the model is detailed in three steps. First, two chains of GANs generate texture and label maps for facades and roofs. Then, the resolution of the generated textures is increased by a dedicated window generation chain and two super-resolution GANs for roofs and facades. Finally, 3D details are generated for roofs, windows, and facades based on the generated textures and label maps.}
    \label{fig:overview}
\end{figure*}

\section{Overview}
\label{sec:overview}

The input to our method is a coarse building mass model with a given set of flat facades and a roof with known hip- and ridge-lines. Our goal is to generate texture and geometric details on the facades and the roof. These details are generated with a cascade of GANs. Our generative approach contrasts with the reconstruction performed by photogrammetric approaches, see Figure~\ref{fig:photogrammetric} for a comparison. In contrast to traditional GAN setups, our method allows control over the style of the synthesized outputs. It generates geometric detail in addition to texture detail. Geometric detail is generated by training the GANs to output an additional \emph{label map} that can be used to select the type of geometry to place at each location on the facade and roof (if any).

We interleave GANs that output structural labels with those that output textures. This leads to several desirable properties that are difficult to obtain with traditional end-to-end GAN setups:
Firstly, the output should exhibit a plausible structure. For example, windows tend to be arranged in grids, and the ground floor usually has a different structure than the other floors. In our experiments we found that training a GAN end-to-end makes it difficult to obtain plausible structure; the structure is never given explicitly as an objective and it must be deduced from the facade texture. The structural labels also allow us to map the output bitmaps to 3D geometry, regularize the outputs using known priors, and permit users to manually modify the structure.
Secondly, the user should have some control over the style of the output. Facades of the same building usually have the same style, while the amount of style variation in a block of buildings depends on the city and area. Generating realistic city blocks therefore requires control over the style. Figure~\ref{fig:photogrammetric} presents a few examples. 
Thirdly, we wish to improve upon the quality achievable with a generic GAN architecture like Pix2Pix~\cite{pix2pix}. While recent work has shown remarkable quality and resolution~\cite{Karras:2018:PGG}, achieving this quality with a single network trained end-to-end comes at a prohibitive resource cost.

We improve these three desirable properties in our outputs at a reasonable resource cost by splitting the traditional single-GAN setup into multiple smaller steps that can be trained and evaluated separately. Synchronization across different steps using a low-dimensional embedding of style ensures the consistency of outputs. Additionally, the style embedding can be manipulated by the user, giving control over the style distribution on a building, a block or multiple blocks.

Figure~\ref{fig:overview} shows an overview of the steps performed to generate facade and roof details. In the following, we assume that the user is working on a block of buildings, but similar steps also apply to multiple building blocks or a single building. First, the user defines style distributions for the building block. Style distributions can be provided for several building \emph{properties}, such as facade texture, roof texture, and window layouts. Each distribution is modeled as a mixture of Gaussians in a low-dimensional \emph{style space}, where the user may choose $n$ modes by providing $n$ reference images (which do not need to be consistent with the building shape or each other), and optionally a custom variance. Each building in the block samples one style vector from this distribution and uses it for all windows, facades and the roof. Specifying styles gives more control over the result, but is optional; the style for any building property can instead be entirely random. Section~\ref{sec:style_control} provides details.

After each building style is defined, two separate chains of GANs with similar architectures generate the facade and roof textures, as well as the corresponding label maps. Each GAN in the chain performs image-to-image mapping based on BicycleGAN~\cite{zhu2017multimodal}. We extend this architecture with several conditional inputs, including a mask of the empty facade or roof and several metrics describing the input, such as its approximate scale and a distance transform of the input boundary. This information about the global context makes it easier for a network that operates only on local patches to make global decisions, such as generating details at the correct scale, or placing objects such as doors at the correct height. Details about the GAN architecture are provided in Section~\ref{sec:gan_architecture}.

To generate facade textures and label maps, three of these GANs are chained together, each performing one step in the construction of the final facade details. The first GAN generates window labels from a blank facade mask; the second GAN transform these labels into the facade texture; and the final GAN detects non-window labels in the facade texture to generate a second detailed label map. Similar steps are performed for roof textures and label maps.
%Section~\ref{sec:detail_generation} provides details.
As we will show, this multi-step approach results in higher-quality details compared to an end-to-end approach.

The resolution of the facade and roof textures is limited by the memory requirements of the GANs. To obtain higher-resolution textures without significantly increasing the memory requirements, we employ two strategies: first, since windows are prominent features that usually exhibit fine details, we texture windows individually. A GAN chain creates window-pane labels in a first step and window textures from these labels in a second step. Second, we increase the resolution of the roof and wall textures using a super-resolution GAN applied to wall patches of fixed size. This GAN has the same architecture as the GANs used in all other steps. Details on all GAN chains are given in Section~\ref{sec:detail_generation}.

Finally,
3D geometric details are created based on the generated label maps using procedural geometry ,
%geometric details are lifted to 3D using procedural geometry based on the generated label maps,
and the resulting detailed mass models are textured using the generated textures.
We also use label maps and textures to define decorative 3D details and material properties maps. Details are provided in Section~\ref{sec:geometry_synthesis}. \changed{The source code and pre-trained network weights are available from the project webpage \emph{\url{http://geometry.cs.ucl.ac.uk/projects/2018/frankengan/}}.}

% single GAN
%

%Input: 3D mass models (flat facade) with dimensions + example facade images for style guidance

%Output: 2.5D geometry; still very low cost; and changeable by different facade images

%facade: blank facade > windows > texture conditioned on window labels > further labels extracted > 2,5 geometry + stylized

%Similarly, 
%roof: flat facade > straight skeleton encoding > roof slopes > textures > dormer/chimney > final textures

%style consistency: inside a floor, on a facade front; across buildings on the same street; using images as guidance

%% file: sections/method.tex
\section{GAN architecture}
\label{sec:gan_architecture}

Our texture and label maps are generated in multiple steps, where each step is an image-to-image transformation, implemented with a GAN. Traditional image-to-image GANs such as Pix2Pix~\cite{pix2pix} can learn a wide range of image-to-image transformations, but the variation of outputs that can be generated for a given input is limited. Since we aim to generate multiple different output styles for a given input, we base our GAN architecture on the recently introduced BicycleGAN architecture~\cite{zhu2017multimodal}, which explicitly encodes the style in a low-dimensional \emph{style space} and allows outputs to be generated with multiple styles. In this section, we briefly recap the BicycleGAN setup and describe our modifications.

Image-to-image GANs train a generator function,
\begin{equation}
\label{eq:gan_function}
B = G(A,Z): \mathbb{R}^n \times \mathbb{R}^k \rightarrow \mathbb{R}^n,    
\end{equation}
that transforms an input image $A$ to an output image $B$. For example, $A$ might be an image containing color-coded facade labels, such as windows, and $B$ a corresponding facade texture. The second input $Z$ is a vector of latent variables that describe properties of the output image that are not given by the input image, such as the wall color. We call $Z$ the \emph{style vector} and the space containing $Z$ the \emph{style space} $\mathcal{Z}$. The embedding of properties into this space is learned by the generator during training. Typically, $Z$ is chosen to be random during training and evaluation, effectively randomizing the style.

The generator's goal is to approximate some desired but unknown joint distribution $p(A,B)$ by training it with known samples from this distribution, in the form of two datasets $\mathcal{A}$ and $\mathcal{B}$ of matching input/output pairs. For example, $p(A,B)$ may be the distribution of matching pairs of facade labels and facade textures. During generator training, the difference between the generated distribution $p(A,G(A,Z))$ and the desired unknown distribution $p(A,B)$ is measured in an adversarial setup, where a \emph{discriminator} function $D$ is trained to distinguish between samples from the two distributions, with the following cross-entropy classification loss:
\begin{equation}
\begin{aligned}
\label{eq:loss_gan_d}
    \mathcal{L}^D_{\textrm{GAN}}(G,D) =\ & \mathbb{E}_{A,B \sim p(A,B)} \bigl[-\log D(A,B)\bigr]\ + \\
    & \mathbb{E}_{A,B \sim p(A,B),\ Z \sim p(Z)}\bigl[-\log \left(1-D(A,G(A,Z))\right)\bigr],
\end{aligned}
\end{equation}
where $p(Z) = \mathcal{N}(0,I)$ is the prior over style vectors, defined to be a standard normal distribution. The generator is trained to output samples that are misclassified by the discriminator as being from the desired distribution:
\begin{align}
\label{eq:loss_gan_g}
    \mathcal{L}^G_{\textrm{GAN}}(G,D) = \mathbb{E}_{A,B \sim p(A,B),\ Z \sim p(Z)}\bigl[-\log D(A,G(A,Z))\bigr].
\end{align}
Additionally, an L1 or L2 loss term between the generated output and the ground truth is usually included:
\begin{align}
\label{eq:loss_gan_l1}
    \mathcal{L}_{\textrm{L1}}(G) = \mathbb{E}_{A,B \sim p(A,B),\ Z \sim p(Z)}\bigl\|B-G(A,Z)\bigr\|_1.
\end{align}

In general, the conditional distribution $p(B|A)$ for a fixed $A$ may be a multi-modal distribution with large variance; for example, there is a wide range of possible facade textures for a given facade label image. However, previous work~\cite{pix2pix} has shown that in typical image-to-image GAN setups, the style vector $Z$ is largely ignored, resulting in a generator output that is almost fully determined by $A$ and restricting $p(B|A)$ to have low variance. To solve this problem, BicycleGAN uses an \emph{encoder} $E$ that obtains the style from an image and combines additional loss terms introduced in previous works~\cite{Donahue:2016:afl,Dumoulin:2016:ali,vae_gan} to ensure that the style is not ignored by the generator.

\begin{figure}[t]
    \centering
    \includegraphics[width=\columnwidth]{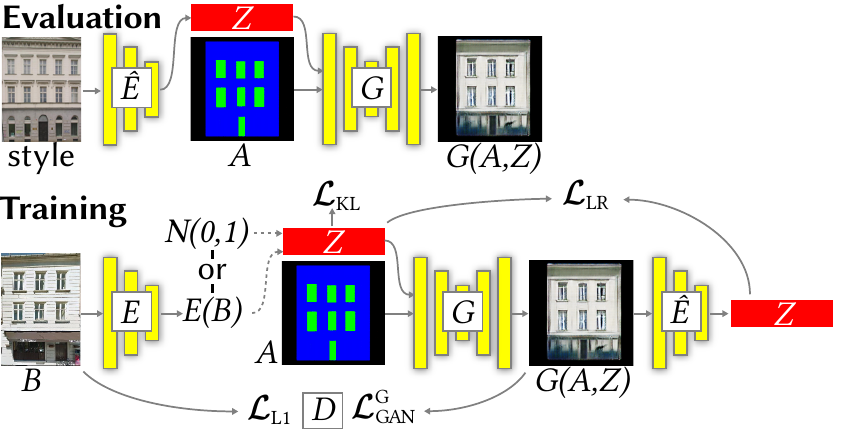}
    \caption{GAN architecture. The setup used during evaluation is shown in the top row, and the training setup is shown in the bottom row. Dotted lines denote random sampling.}
    \label{fig:gan}
\end{figure}

First, based on ideas from Variational Autoencoders~\cite{vae}, the encoder outputs a distribution $E(B)$ of styles for each image instead of a single style vector. In Equations~\ref{eq:loss_gan_d} to~\ref{eq:loss_gan_l1}, $p(Z) = E(B)$ is used instead of the standard normal distribution. The distribution $E(B)$ is regularized to be close to a standard normal distribution to encourage style vectors to form a large contiguous region in style space that can easily be sampled:
\begin{align}
\label{eq:loss_gan_kl}
    \mathcal{L}_{\textrm{KL}}(E) = \mathbb{E}_{B \sim p(B)}\big[\mathcal{D}_{KL}(E(B) \| \mathcal{N}(0,I))\bigr],
\end{align}
where $\mathcal{D}_{KL}$ is the KL-divergence.
Second, the generator is encouraged not to ignore style by including a style reconstruction term:
\begin{align}
\label{eq:loss_gan_lr}
    \mathcal{L}_{\textrm{LR}}(E) = \mathbb{E}_{A \sim p(A), Z \sim \mathcal{N}(0,I)}\big\|Z - \hat{E}(G(A,Z))\bigr\|_1,
\end{align}
where $\hat{E}$ denotes the mean of the distribution output by $E$.
Intuitively, this term measures the reconstruction error between the style given to the generator as input and the style obtained from the generated image. The full loss for the generator and encoder is then:
\begin{equation}
\begin{aligned}
\label{eq:loss_gan}
    \mathcal{L}^G(G,D,E) =\ &
    \lambda_{\textrm{GAN}}\mathcal{L}^G_{\textrm{GAN}}(G,D) +
    \lambda_{\textrm{L1}}\mathcal{L}_{\textrm{L1}}(G)\ + \\
    & \lambda_{\textrm{KL}}\mathcal{L}_{\textrm{KL}}(E) +
    \lambda_{\textrm{LR}}\mathcal{L}_{\textrm{LR}}(E).
\end{aligned}
\end{equation}
The hyper-parameters $\lambda$ control the relative weight of each loss. A diagram of this architecture is shown in Figure~\ref{fig:gan}.

\systemName trains a BicycleGAN for each individual step, with one exception that we discuss in Section~\ref{sec:franken_gan}. In addition to style, we also need to encode the real-world scale of the input mass model, so that the output details can be generated in the desired scale.
%
%We augment BicycleGAN's inputs with additional conditions; the first input is output scale so that the network can match the world scale of a given mass model. This conditional channel is appended to $A$.
%
%In addition to style control, control of the output scale is also needed to match the world scale of a given mass model
We condition the GANs on an additional constant input channel that contains the scale of the facade in real-world units. This channel is appended to $A$.

\begin{figure}[t]
    \centering
    \includegraphics[width=\columnwidth]{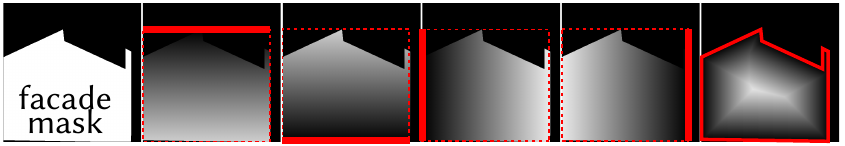}
    \caption{Additional input channels. GANs are conditioned on additional channels that include information about the global context at each pixel. Given a facade/roof mask, we include the distance to the facade boundary and the distance to each bounding box side, making it easy for the network to decide how far it is from the boundary and at what height on the facade.}
    \vspace{-5pt}
    \label{fig:context_info}
\end{figure}

In our experiments, we observed that using this BicycleGAN setup to go directly from inputs to outputs in each step often resulted in implausible global structure, such as misaligned windows or ledges on facades. The limited receptive field of outputs in both the generator and the discriminator constrains coordination between distant output pixels, making it difficult to create globally consistent structure. Increasing the depth of the networks to increase the receptive field alleviates the problem but has a significant resource cost and destabilizes training. We found that conditioning the GANs on additional information about the global context of each pixel was more efficient. More specifically, we conditioned the GANs on five additional channels that are appended to $A$: the distance in real world units to each side of the bounding box and to the nearest boundary of a facade or roof. Examples are shown in Figure~\ref{fig:context_info}.

\begin{figure*}[t!]
    \centering
    \includegraphics[width=\textwidth]{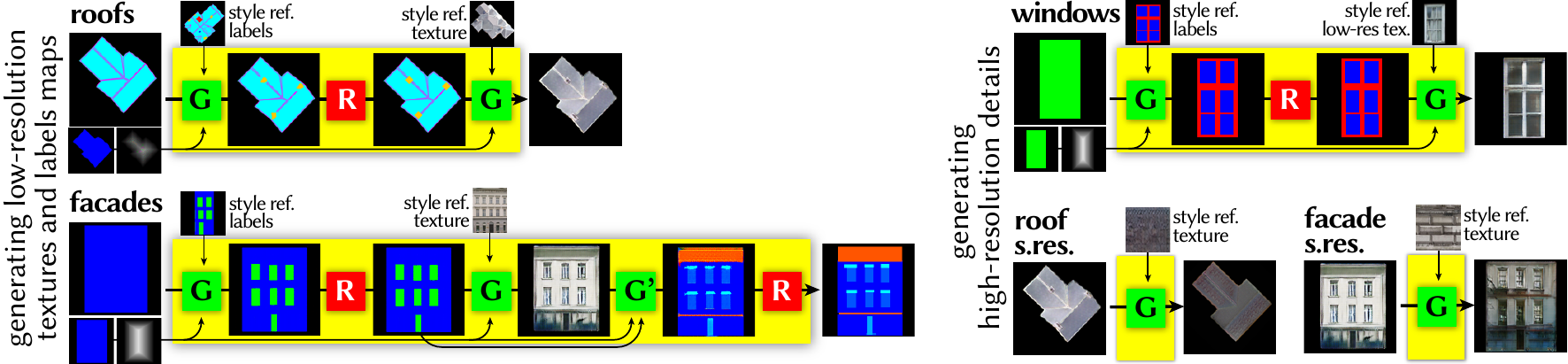}
    \caption{\systemName details. Each GAN chain (yellow rectangles) consists of several GANs (G) that each perform an image-to-image transformation. GANs are usually conditioned on additional inputs (arrows along the bottom) and are guided by a reference style (arrows along the top). Label outputs are regularized (R) to obtain clean label rectangles. Figure~\ref{fig:overview} shows these chains in context.}
    \label{fig:franken_gan}
\end{figure*}

\begin{figure}[b]
    \centering
    \includegraphics[width=\columnwidth]{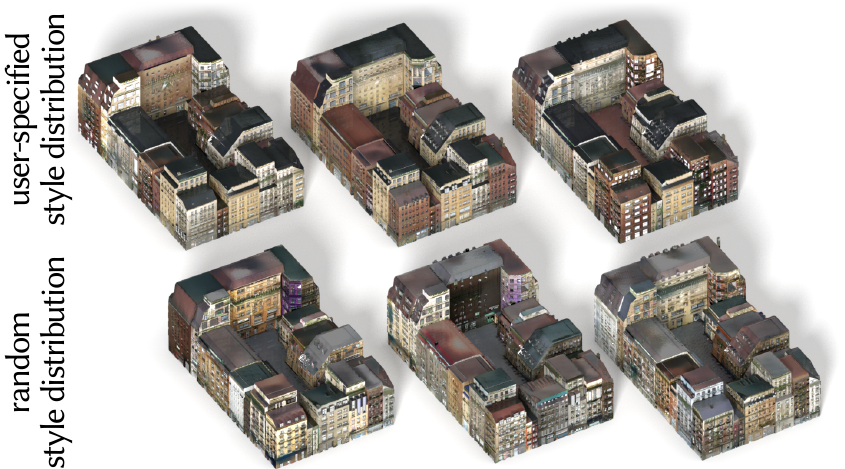}
    \caption{Specifying a style distribution gives control over the generated details. The top row shows three results generated with the same user-specified style distribution, while the bottom row uses the style prior, giving random styles. Note how the buildings in the top row have a consistent style while still allowing for some variation (depending on the variance chosen by the user), while the bottom row does not have a consistent style.}
    \label{fig:specified_vs_random_style}
\end{figure}

\section{FrankenGAN}
\label{sec:franken_gan}

Detail generation is performed by a cascade of textures and label maps, as shown in Figure~\ref{fig:overview}. These are generated by \systemName in several separate chains of GANs, where each GAN is trained and run independently. Splitting up this task into multiple steps rather than training end-to-end has several advantages. First, we can provide more guidance in the form of intermediate objectives, for example window layouts, or low-resolution textures. In our experiments, we show that such intermediate objectives provide a significant advantage over omitting this guidance. While in theory there are several ways to provide intermediate objectives for end-to-end networks, for example by concatenating our current GANs, this would result in extremely large networks, leading to the second advantage of our approach: GAN training is notoriously unstable, and training small GANs is more feasible than training large ones. An end-to-end network with intermediate objectives would need to have a very large generator with multiple discriminators, making stable training difficult achieve. In addition, splitting the network reduces resource costs during training. Instead of a single very large network, we can separately train multiple smaller networks. Note that training a very large network one part at a time
%only keeping part of a very large network in memory during training
%is an option in theory, but in practice it
would require storing and loading parts of the network from disk in each forward and backward pass, which is prohibitive in terms of training times. Finally, using separate GANs, we can regularize intermediate results with operations that are not are not differentiable or would not provide a good gradient signal.

\subsection{Style Control}
\label{sec:style_control}

One difficulty with using separate GANs is achieving stylistically consistent results. For example, windows on the same facade usually have a similar style, as do ledges or window sills. Style control is also necessary beyond single roof or facade textures: adjacent facades on a building usually look similar, and city blocks may have buildings with similar facade and roof styles. A comparison of generated details with and without style control is given in Figure~\ref{fig:specified_vs_random_style}.
In \systemName, style can be specified for eight properties: the coarse facade and roof texture, facade and roof texture details, such as brick patterns, the window layout on the facade, the glass pane layout in the windows, the window texture, and the layout of chimneys and windows on a roof. The user can describe the style distribution of a property over a building block with a mixture of isotropic Gaussians in style space $\mathcal{Z}$:
\begin{equation}
    p(Z|\mathbf{S}, \bm{\sigma}) = \sum _{i=1}^{m}\phi_{i}\mathcal{N}(E(S_i),\sigma_{i}),
\end{equation}
where $Z \in \mathcal{Z}$ is the style vector, $\mathcal{N}$ is the normal distribution and the weights $\phi_{i}$ must sum to $1$. $Z$ provides a compact representation of style. We use an eight-dimensional style space in our experiments. The means of the Gaussians are specified by encoding $m$ style reference images $S_i \in \mathbf{S}$ with the encoder described in the previous section. The variance $\bm{\sigma} = (\sigma_1, \dots, \sigma_m)$ specifies the diversity of the generated details and can be adjusted per reference image with a slider. One of these distributions can be specified per property and the styles are sampled independently.

In many cases, however, the styles of different properties are dependent. For example, the color of roofs and facades may be correlated. To specify these dependencies, several sets of property distributions may be specified, each set $\mathcal{S}_i = \{(\mathbf{S}_p, \bm{\sigma}_p)\}_{p=1 \dots 8}$ contains one mixture model per property $p$. For each building, one of these sets is chosen at random. The special case of having a single Gaussian ($m=1$) per property in each set effectively gives a Gaussian mixture model over the \emph{joint} space of all properties, with each set being one component. The user does not need to provide the style for all properties. Any number of properties may be left unspecified, in which case the style vector is sampled from the GAN's style prior, which is a standard normal distribution.

\subsection{Detail Generation}
\label{sec:detail_generation}
\systemName uses five chains of GANs, which can be split into two groups: two chains for generating initial coarse details (textures and label maps) for roofs and facades, and three chains for increasing the resolution given the coarse outputs of the first group. Details of these chains are shown in Figure~\ref{fig:franken_gan}. Most of the chains have intermediate results, which are used for the geometry synthesis that we will describe in Section~\ref{sec:geometry_synthesis}. Each GAN takes an input image and outputs a transformed image. In addition to the input image, all GANs except for the super-resolution networks are conditioned on the scale and context information described in the previous section, making it easier to generate consistent global structure. Each GAN is also guided by a style that is drawn from a distribution, as described above. Figure~\ref{fig:franken_gan} shows reference images $S_i$ that are used to specify the distribution. Images output by GANs are either label maps $L$, or textures $T$. Each label map output by a GAN is passed through a regularizer $R$, denoted by the red rectangles in Figure~\ref{fig:franken_gan}, to produce a clean set of boxes $R(L)$ before being passed to the next step. We now describe each chain in detail. The regularizers are described in Section~\ref{sec:regularizers}.
%For brevity, we will not explicitly describe scale, context and style in each chain, and assume they are included implicitly for each GAN.

%$L_r$ %$T_r$ $L^{\sim}_r$ $L_r = (R_{Lr} \circ G_{Lr})(L^{\sim}_r)$ $T_r = G_{Tr}(L_r)$
\paragraph{Roofs}
The roof chain generates roof detail labels and the coarse roof texture. The chain starts with a coarse label map  of the roof as the input image. This label map includes ridge and valley lines of the roof, which are obtained directly from the mass model. Flat roofs are labeled with a separate color. The first GAN adds chimneys and pitched roof windows. These labels are regularized and then used by the second GAN to generate the coarse roof texture.

\begin{figure}[t]
    \centering
    \includegraphics[width=\columnwidth]{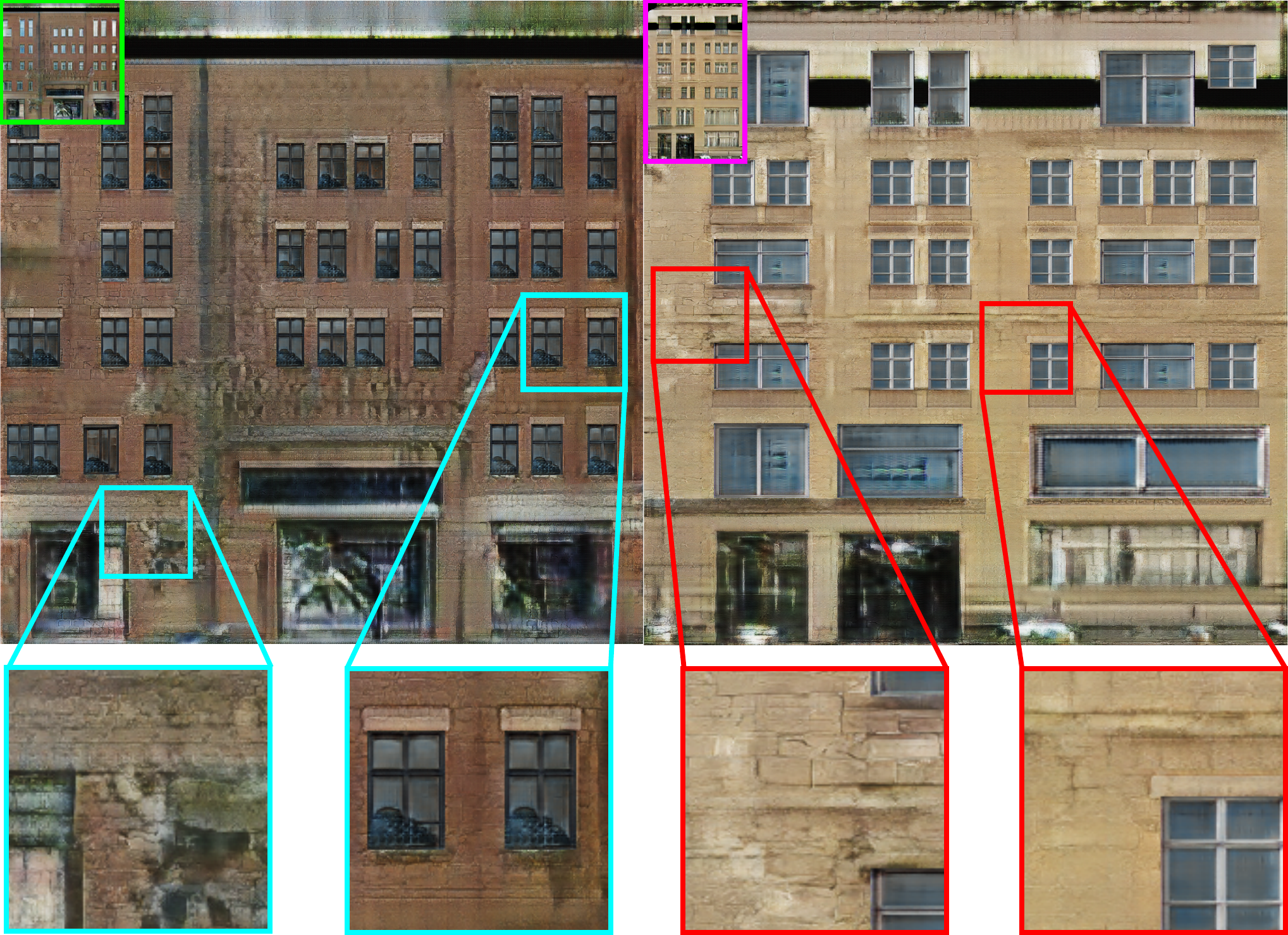}
    \caption{Super-resolution. Given inputs (green, magenta), the super-resolution network creates high-quality textures for the walls, while the window GAN chain provides high-quality windows. Note that the window label and window texture networks each run once for every window.}
    \vspace{-5pt}
    \label{fig:super}
\end{figure}

\paragraph{Facades}
The facade chain generates window labels, full facade labels, and the coarse facade texture. Window labels are generated separately from the full labels, since they may be occluded by some of the other labels. The first GAN starts by creating window and door labels from facade boundaries. These are regularized before being used as input in the second GAN to generate the coarse facade texture. The third GAN detects the full set of facade labels from the facade texture, including ledges, window sills, and balconies, which are also regularized to give a cleaner set of labels. The third GAN has a different architecture: since we expect there to be a single correct label map for each facade texture, we do not need style input, which simplifies the GAN to the Pix2Pix architecture~\cite{pix2pix}. Since the window and door labels are known at this point, we also condition this GAN on these labels. Detecting the full label set from the facade texture instead of generating it beforehand and using it as input for the texture generation step is a design choice that we made after experimenting with both variants. Detail outlines in the generated texture tend to follow the input labels very closely, and constraining the details in this way results in unrealistic shapes and reduced variability. For all three GANs, areas occluded by nearby facades are set to the background colour; this ensures that the feature distribution takes into account the nearby geometry.

Since the layout of dormer windows needs to be consistent with the facade layout, we create these windows in the facade chain. More specifically, we extend the input facade mask with a projection of the roof to the facade plane. This allows us to treat the roof as part of the facade and to generate a window layout that extends to the roof. Roof features (chimneys or pitched windows) that intersect dormer windows are removed. 

\paragraph{Windows}
To obtain high-resolution window textures, we apply the window chain to each window separately, using a consistent style. Each window is cut out from the window label map that was generated in the facade chain, and scaled up to the input resolution of the GAN. The steps in the window chain are then similar to those in the roof chain. We generate glass pane labels from the window region, regularize them, and generate the high-resolution window texture from the glass pane labels. %Different from the roof chain, the style of the window texture GAN is not sampled from a given distribution, but is instead encoded from the window patch in the coarse facade texture to maintain consistency with the surrounding facade.
%\paul{do we still use the low-res window texture as style, or is it manually defined?}
% If manual, this needs to be changed here, number of properties increased by 1, and the overview figure should contain an example of a window texture style

\paragraph{Super-resolution}
High-resolution roof and facade textures are obtained with two GANs that are trained to generate texture detail, such as bricks or roof tiles from a given low-resolution input. Roof and facade textures are split into a set of patches that are processed separately. Each patch is scaled up to the input size of the GANs before generating the high-resolution output. Consistency can be maintained by fixing the style for the entire building. The output patches are then assembled to obtain the high-resolution roof and facade textures. Boundaries between patches are blended linearly to avoid seams. Examples are shown in Figure.~\ref{fig:super}. Our interleaved GANs allow us to augment the super-resolution texture map with texture cues from the label maps. For example, window sills are lightened, and roof crests are drawn; these augmentations take the form of drawing the labels in a single colour with a certain alpha. 
\emph{Note that because of the large scale of the super-resolution bitmaps, we explicitly state which figures use these two networks.}

\subsection{Regularizers}
\label{sec:regularizers}
Our GANs are good at producing varied label maps that follow the data distribution in our training set. Alignment between individual elements is, however, usually not perfect. For example, window sills may not be rectangular or have different sizes in adjacent windows, or ledges may not be perfectly straight. Although the discrepancy is usually small, it is still noticeable in the final output. Our multi-step approach allows us to use any non-differentiable (or otherwise) regularization. We exploit domain-specific knowledge to craft simple algorithms to improve the alignment of the label maps. We then provide 3D locations for geometric features. In the following, we describe our regularizers in detail.

\begin{figure}[t!]
    \centering
    \includegraphics[width=\columnwidth]{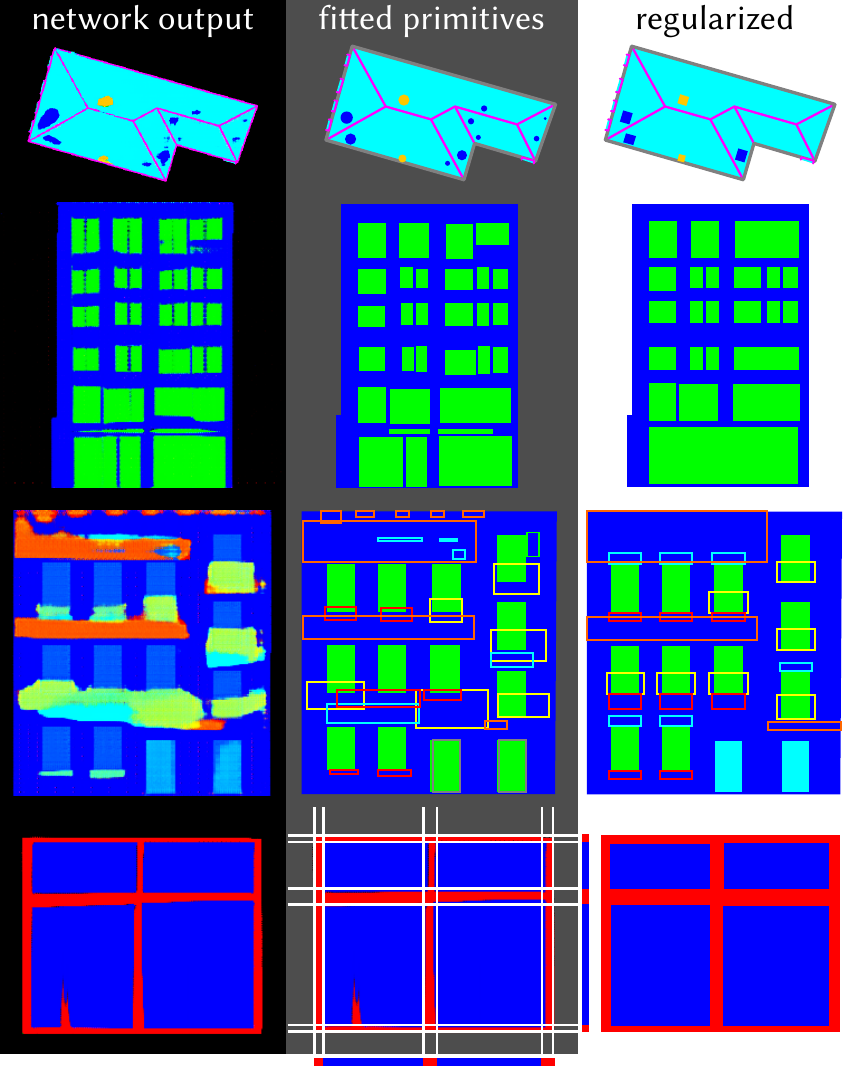}
    \caption{\changed{Regularization. All our label maps are regularized before being used by the next GAN in the chain.
    %Columns: network outputs (images), fitted primitives (boxes and circles), and regularized results.
    Rows: roof details, facade windows, facade details, and window details (regularized with an outer product). }}
    \vspace{-5pt}
    \label{fig:regularizers}
\end{figure}

%\begin{figure}[t!]
%    \centering
%    \includegraphics[width=\columnwidth]{images/regularizers.pdf}
%    \caption{Window detail regularization. All our label maps are regularized before being used by the next GAN in the chain. Here, we regularize window glass pane labels using the outer product of two 1D label masks.}
%    \label{fig:regularizers}
%\end{figure}

\paragraph{Roof detail labels}
\changed{Chimneys and pitched roof windows are regularized by fitting circles to the network output. These are then converted to rectangles, which may be oriented to the roof pitch, see Figure~\ref{fig:regularizers}, top row. We take the center of each connected component of a given label and use its area to estimate a circle size. Smaller circles are removed before we convert each to a square. We observe that roof features such as chimneys and roof windows are typically aligned to the roof's slope. We therefore orient the bottom edge of each feature to the gutter of the associated roof pitch.  Finally, we shrink it so that it lies entirely within a single roof pitch and avoids dormer windows.}

\paragraph{Facade window and door labels}
The window and door layout on a facade has to be regularized without removing desirable irregularities introduced by the GAN that reflect the actual data distribution, such as different window sizes on different floors, or multiple overlayed grid layouts. We start by fitting axis-aligned bounding boxes to doors and windows \changed{(see Figure~\ref{fig:regularizers}, second row center). Then we collect a set of properties for each window, including the x and y extents and the spacing between neighbours, over which we perform a mean-shift clustering. We use a square kernel of size 0.4 meters for each property, until convergence or a maximum of 50 mean-shift iterations. This ensures that these properties can have a multi-modal distribution, which preserves desirable irregularities, while also removing small-scale irregularities (see Figure~\ref{fig:regularizers}, second row right).}

\paragraph{Facade detail labels}
Since adjacent details are often not perfectly aligned, we snap nearby details, such as window sills and windows to improve the alignment. We also observed that in the generated label map, the placement of small details such as window sills and moldings is sometimes not coordinated over larger distances on the facade. To improve regularity, we propagate details such as window sills and moldings that are present in more than $50\%$ of the windows in a row to all remaining windows \changed{(see Figure~\ref{fig:regularizers}, third row).}

\paragraph{Window detail labels}
The glass pane layout in a window is usually more regular than the window layout on facades, allowing for a simpler regularization: we transform the label map into a binary glass pane mask and approximate this 2D mask by the outer product of two 1D masks, one for the columns, and one for the rows of the 2D mask. This representation ensures that the mask contains a grid of square glass panes. The two 1D masks are created by taking the mean of the 2D mask in the x- and y-directions, and thresholding them at $0.33$. An example is shown in Figure~\ref{fig:regularizers}, bottom row.

\begin{figure}[t!]
    \centering
    \includegraphics[width=\columnwidth]{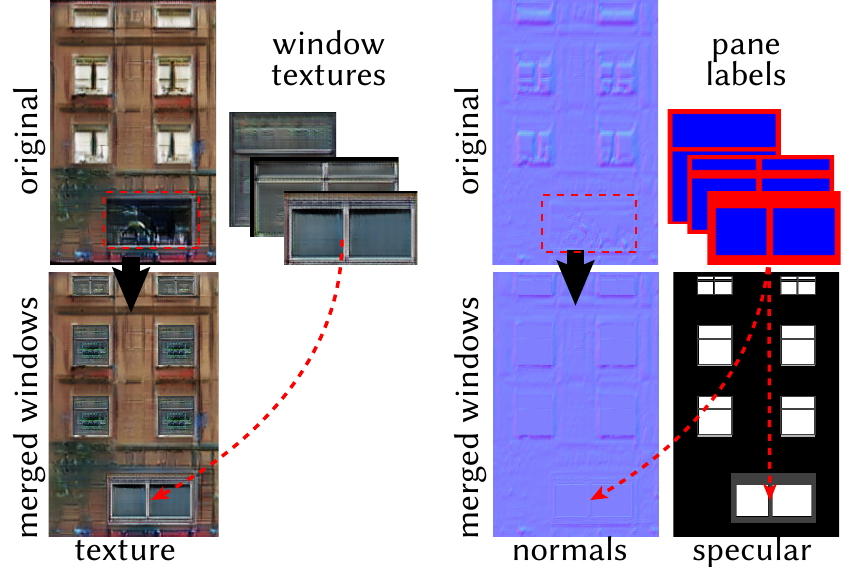}
    \caption{Generated window textures and labels are merged back into the facade texture, increasing fidelity of window textures, normals and materials.}
    \label{fig:merge_maps}
\end{figure}

% \subsection{Facade Textures}
% \label{sec:facade_textures}
% \subsection{Roof Textures}
% \label{sec:roof_textures}
% \subsection{Window Textures}
% \label{sec:window_textures}
% \subsection{Facade Super-resolution}
% \label{sec:super_resolution}

\subsection{Geometry Synthesis}
\label{sec:geometry_synthesis}

As output of the five GAN chains, we have high-resolution roof, facade, and window textures, as well as regularized label maps for roof details, facade details, and window panes. These are used to generate the detailed mass model. First, geometry for details is generated procedurally based on the label maps. For details such as window sills and ledges, we use simple extrusions, while balconies and chimneys are generated with small procedural programs to fit the shape given by the label map. 

To apply the generated textures to the detailed mass models, UV maps are generated procedurally along with the geometry. In addition to textures, we also define building materials based on the label maps. Each label is given a set of material properties: windows, for example, are reflective and have high glossiness, while walls are mostly diffuse. To further increase the fidelity of our models, textures and label maps are used to heuristically generate normal maps. The intensity of the generated texture is treated as a height field that allows us to compute normals. While this does not give us accurate normals, it works well in practice to simulate the roughness of a texture. Per-label roughness weights ensure that details such as glass panes still remain flat. Finally, generated window textures and label maps are merged back into the facade textures; an example is given in Figure~\ref{fig:merge_maps}.

\begin{figure*}[t]
    \centering
    \includegraphics[width=\textwidth]{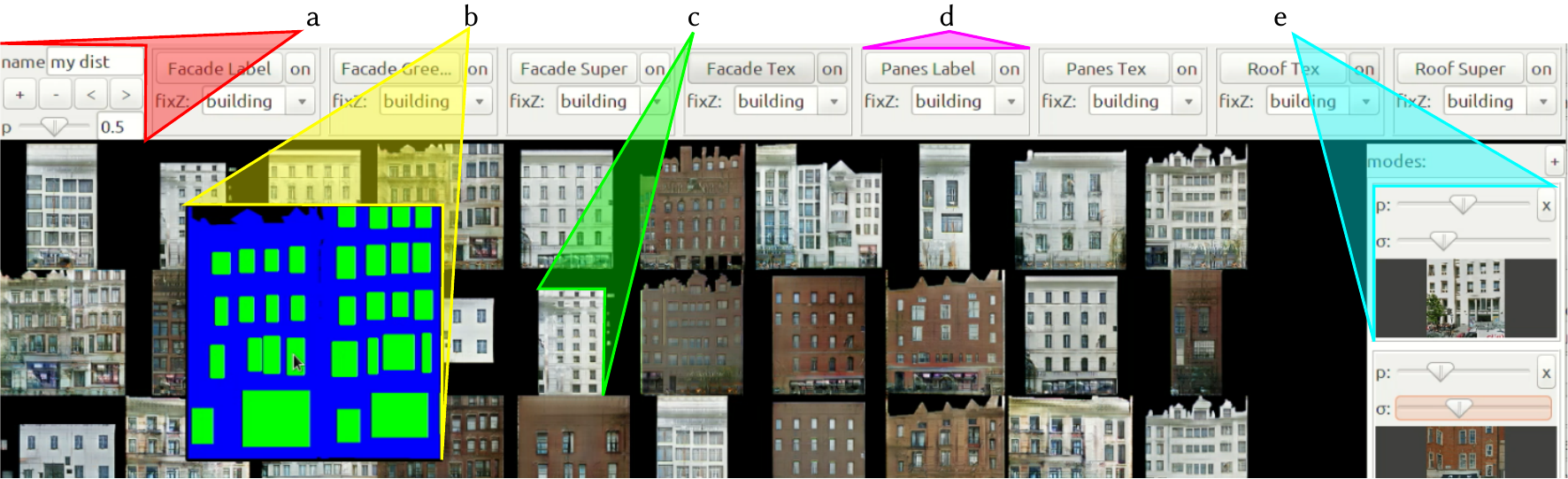}
    \caption{The distribution designer UI (see supplemental video). Given a style distribution (a), the system continuously shows evaluations of that distribution (c). By clicking on an image, the user can see the network inputs (b). Different networks can be selected (d). The style distribution for any network is a Gaussian mixture model that may have multiple modes (e), the mean of which is given by an exemplar image.}
    \vspace{-5pt}
    \label{fig:ui}
\end{figure*}

\begin{figure*}[t]
    \centering
    \includegraphics[width=\textwidth]{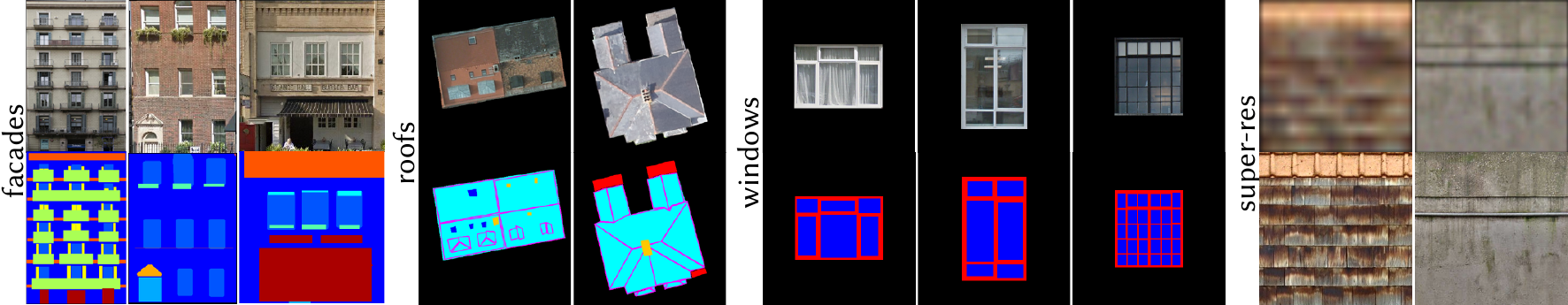}
    \caption{Datasets used to train our GANs. We use four datasets of labeled images, a few examples are shown here.}
    \label{fig:dataset}
\end{figure*}

\subsection{User interface}

Our system contains a complete framework for interactively using \systemName. A user may select an urban city block, specify a distribution, and then the system adds the additional geometry and textures to the 3D view. At this point, the user can edit semantic details (such as window locations), while seeing texture updates in real time. Of note is our interface to build our joint distributions (Figure~\ref{fig:ui}), which continually shows the user new examples drawn from the current distribution. The accompanying video demonstrates the user interface.
%\changed{\sout{The source code of our system, and weights for accompanying networks, will be made available online.}}

%% file: sections/results.tex
\section{Results}
\label{sec:results}
We evaluate our method on several scenes consisting of procedurally generated mass models. 
%created with procedural extrusions~\cite{Kelly:SIGA:2017}.
We qualitatively show the fidelity of our output and the effects of style and scale control. Style and scale control are also evaluated quantitatively with a perceptual study, and we provide comparisons to existing end-to-end networks, both qualitatively and quantitatively through another perceptual study. 

\begin{figure*}[t!]
    \centering
    \includegraphics[width=\textwidth]{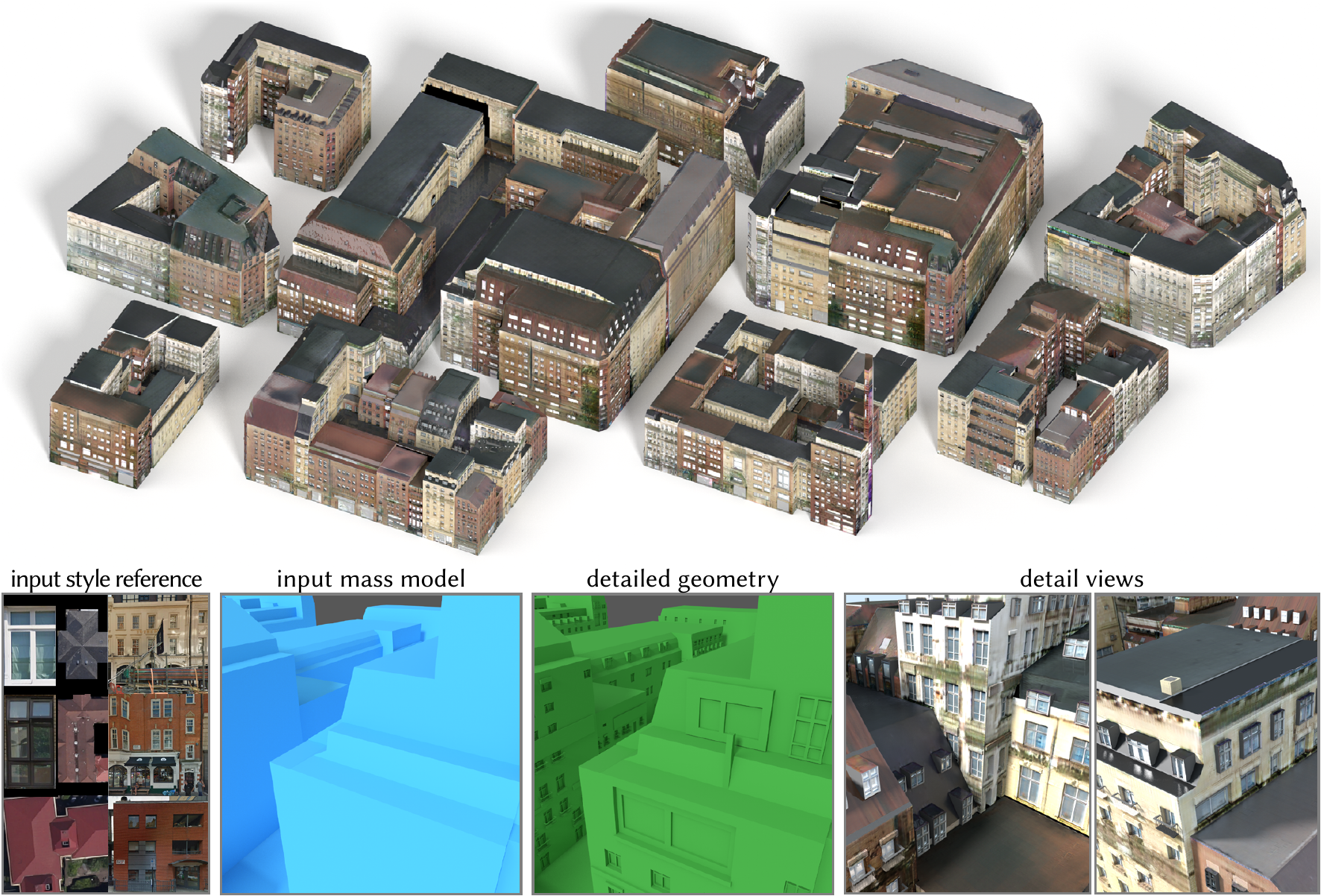}
    \caption{Detailed London area. The output of our method is shown on top, and the input style images at the bottom left. This is followed, from left to right, by close-ups of the input mass models, detail geometry generated by our method, and two detailed views of the generated model using super-resolution textures.}
    \label{fig:london}
\end{figure*}

\subsection{Datasets and Training Setup}

Each GAN in our framework performs an image-to-image transformation that is trained with a separate dataset of matched image pairs. Matched images for these pairs are obtained from three datasets:

The \emph{facade} dataset consists of the CMP dataset~\cite{cmp_dataset} and a larger dataset of labeled facades that has not yet been released, but that has been made available to us by the authors. The combined dataset contains $3941$ rectified facades with labels for several types of details, including doors, windows, window sills, and balconies.
We further refined this dataset by removing heavily occluded facades and by annotating the height of a typical floor in each facade to obtain the real-world scale that our GANs are conditioned on, as described in Section~\ref{sec:gan_architecture}.
From this dataset, we create matched pairs of images for each GAN in the facade chain.
%: (facade mask, window layout)-pairs, (window layout, facade texture)-pairs, and (facade texture, full facade label)-pairs.

The \emph{roof} dataset consists of $585$ high-quality roof images with labeled roof area, ridge/valley lines, pitched windows and chimneys. The images are part of an unreleased dataset. We contracted professional labellers to create high-quality labels.
From this dataset, we created matched pairs of images for the two GANs in the roof chain.
%: (coarse roof label, detailed roof label)-pairs and (detailed roof label, roof texture)-pairs.

The \emph{window} dataset contains $1376$ rectified window images with labeled window areas and glass panes. These images were obtained from Google Street View, and high-quality labels were created by professional labellers.
From this dataset, we created matched pairs of images for the two GANs in the window chain. Examples from all datasets are shown in Figure~\ref{fig:dataset}.

In addition to these three datasets, the super-resolution GANs were trained with two separate datasets. These datasets were created from a set of high-quality roof/wall texture patches that was downloaded from the internet, for example, brick patterns or roof shingles, and blurring them by random amounts. The networks are then trained to transform the blurred image to the original image. We found that it increased the performance of our networks to add in a second texture at a random location in the image. This accounts for changes in the texture over a facade or roof that occur in natural images.

\begin{table}
\centering
\small

\caption{GAN statistics:
%Summary of \textsc{FrankenGANs};
the size of the training data (n), resolution (in pixels squared), number of epochs trained, and if the network takes style as input.}
%\multicolumn{5}{c}{\textsc{London}} \\
%\hline
%block & roofs & facades & windows & time (s) \\
%\hline

\begin{tabular}{ccccc}
\hline
& n & resolution & epochs & style \\
\hline
roof labels & 555 & 512 & 400 & yes \\
roof textures & 555 & 512 & 400 & yes \\
facade window labels & 3441 & 256 & 400 & yes \\
facade textures & 3441 & 256 & 150 & yes \\
facade full labels & 3441 & 256 & 335 & no \\
window labels & 1176 & 256 & 200 & yes \\
window textures & 1176 & 256 & 400 & yes \\
facade super-resolution & 2015 & 256 & 600 & yes \\
roof super-resolution & 1122 & 256 & 600 & yes \\
\hline
\end{tabular}

\label{table:net_statistics}
\end{table}

To train each GAN, we alternate between discriminator optimization steps that minimize Eq.~\ref{eq:loss_gan_d} and generator/encoder optimization steps that minimize Eq.~\ref{eq:loss_gan}. The optimization is performed with Adam~\cite{adam}. The weights $(\lambda_{\textrm{GAN}}, \lambda_{\textrm{L1}}, \lambda_{\textrm{KL}}, \lambda_{\textrm{LR}})$ in Equation~\ref{eq:loss_gan} are set to $(1, 10, 0.01, 0.5)$. A large $\lambda_{\textrm{L1}}$ encourages results that are close to the average over all training images. This helps to stabilize training for textures. For label map generation, however, there is usually one dominant label, such as the wall or the roof label, and a large $L1$ loss encourages the generation of this label over other labels. Lowering the $L1$ loss to $1$ improves results for label maps.
% also update GAN architecture section to include the final sum of energy terms with weights
% and maybe discuss the tradeoff between l1 weight and discriminator there? (that L1 should be small for label map generation), but probably better only here
% unbias L1 weight for:
% - image2celabels
% lower L1 weight for:
% - w3_empty2labels
% - r3_clabels2labels
% - empty2windows
Statistics for our GANs are summarized in Table~\ref{table:net_statistics}.

At test time, we evaluate our networks on mass models created through procedural reconstruction from photogrammetric\linebreak meshes~\cite{Kelly:SIGA:2017}. Facades and roofs on these mass models are labeled, and roof ridges and valleys are known, providing all necessary inputs for \systemName: facade masks and coarse roof label maps. Note that it obtaining these inputs automatically from any type of reasonably clean mass model, is also feasible.

\begin{figure*}[t!]
    \centering
    \includegraphics[width=\textwidth]{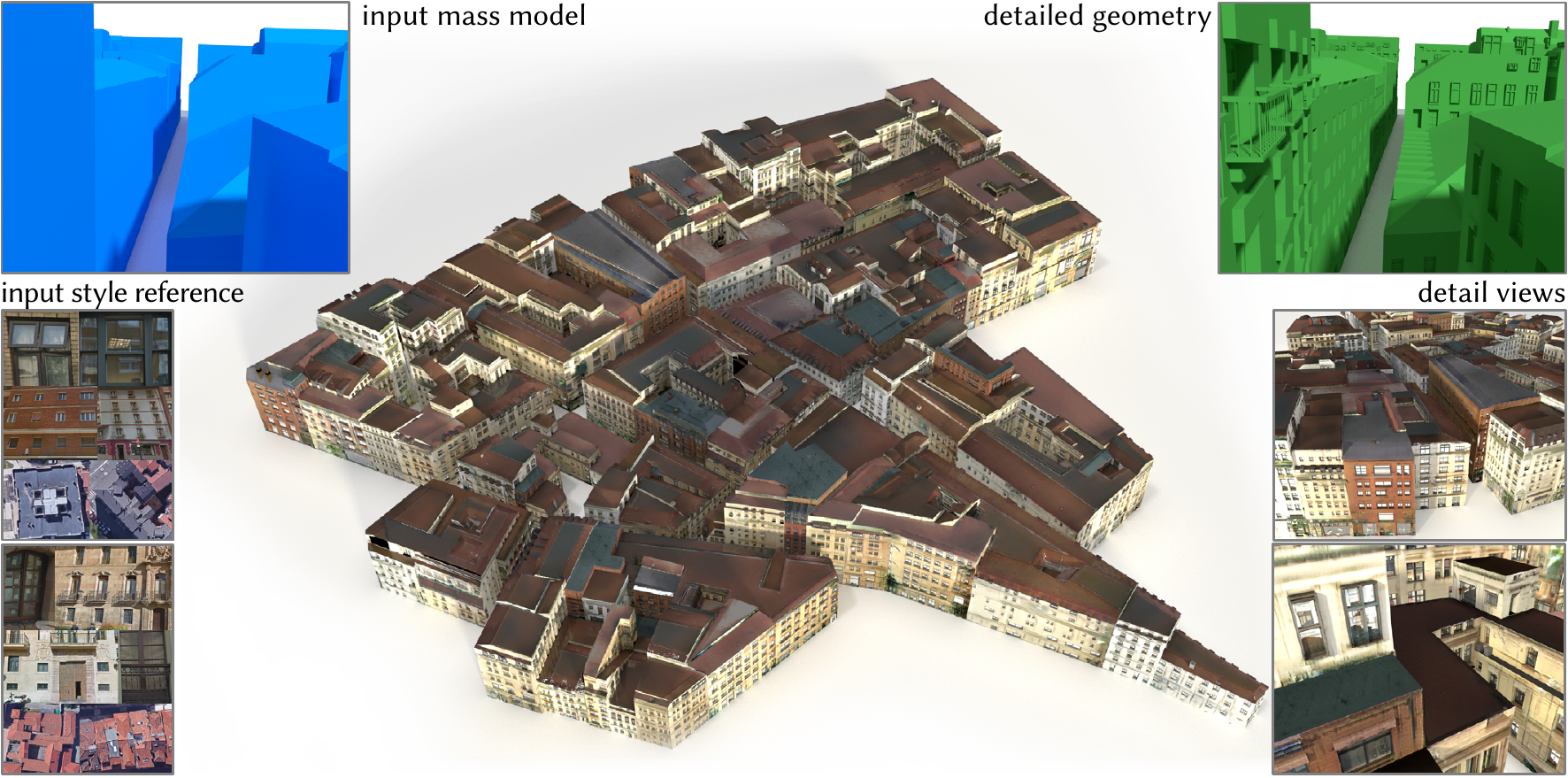}
    \caption{Detailed Madrid area. Input style images and mass models are shown on the left, an overview of our output in the center, and close-ups of our output and the generated geometry (green), showing details like balconies and window moldings, on the right. Lower right panel uses  super-resolution textures.}
    \label{fig:madrid}
\end{figure*}

\subsection{Qualitative Results}

We show the quality of our results with one area of Madrid and one area of London, both spanning several blocks. As our reference style, we take images of facades, windows and roofs, some of which are taken from our dataset, and some from the web. Figures~\ref{fig:teaser} and \ref{fig:london} show the result of detailing the London scene. Reference style textures and input mass models are shown on the left, our detailed result on the right. We produce varied details that are not completely random, but guided by the style given as input. In this example, several sets of style textures are used. Note the varied window layouts and textures: each building has unique layouts and textures that are never repeated. 

Figure~\ref{fig:madrid} shows our results of detailing the Madrid scene. Reference style images and a detailed view of the input mass models are shown on the left; our output is shown in the center and on the right, including a detailed view of the generated geometry. Note several modes of matching styles in the result, including red buildings with black roofs and yellow buildings that often have several ledges.

\begin{table}
\centering

\setlength{\tabcolsep}{1.5pt}

\caption{Statistics for the London and Madrid scenes. We show the number of roofs, facades and windows in each block, as well as the time taken to generate the block.}
\footnotesize
\begin{minipage}{.49\columnwidth}
\begin{tabular}{cccccc}
\multicolumn{5}{c}{\textsc{London}} \\
\hline
block & roofs & facades & windows & time(s) \\
\hline
1 & 29 & 145 & 1075 & 490 \\
2 & 20 & 204 & 541 & 351 \\
3 & 15 & 87 & 536 & 222 \\
4 & 25 & 133 & 1040 & 400 \\
5 & 47 & 243 & 2107 & 809 \\
6 & 27 & 171 & 1622 & 597 \\
7 & 10 & 65 & 1144 & 403 \\
8 & 7 & 40 & 559 & 199 \\
9 & 8 & 42 & 786 & 271 \\
10 & 26 & 158 & 1566 & 577 \\
\hline
\textbf{total} & 214 & 1288 & 10976 & 4322 \\
\hline
\end{tabular} 
\end{minipage}
\hfill
%\hspace{1pt}
%
\begin{minipage}{.49\columnwidth}
\begin{tabular}{cccccc}
%\multicolumn{5}{c}{ } \\
\multicolumn{5}{c}{\textsc{Madrid}} \\
\hline
block & roofs & facades & windows & time(s) \\
\hline
1 & 22 & 146 & 773 & 315 \\
2 & 20 & 110 & 559 & 239 \\
3 & 17 & 103 & 471 & 219 \\
4 & 12 & 67 & 399 & 166 \\
5 & 7 & 66 & 230 & 102 \\
6 & 22 & 116 & 758 & 291 \\
7 & 25 & 139 & 571 & 292 \\
8 & 22 & 125 & 611 & 255 \\
9 & 35 & 240 & 1219 & 495 \\
10 & 37 & 222 & 738 & 350 \\
\hline
\textbf{total} & 219 & 1334 & 6329 & 2722 \\
\hline
\end{tabular}
\end{minipage}

\vspace{-5pt}

\label{table:scene_statistics}
\end{table}

Statistics for the London and Madrid scenes are shown in Table~\ref{table:scene_statistics}. Each of the scenes has $10$ blocks and contains an average of $21$ buildings (the number of buildings equals the number of roofs). All of the generated details, including their textures, are unique in the entire scene. In the London scene, we generate approximately $860$ Megapixels of texture and $2.68$ million triangles; in the Madrid scene we generate approximately $560$ Megapixels and $1.17$ million triangles. The time taken to generate the scenes on a standard desktop PC with an Intel 7700k processor, 32 GB of main memory, and an NVidia GTX 1070 GPU is shown in the last column of Table~\ref{table:scene_statistics}. The \systemName implementation has two modes - when minimizing GPU memory, a single network is loaded at a time, and all applicable tasks in a scene are batch processed; when GPU memory is not a constraint, the whole network can be loaded at once. These modes use 560 MB and 2371 MB of GPU memory, respectively, leaving sufficient space on the graphics card to interactively display the textures. These figures compare favourably with larger end-to-end networks that require more memory~\cite{Karras:2018:PGG}.

%Figure \ref{fig:teaser} has 1.78 million verts, 2.68m triangles. 
%It has 7306 bitmaps 256x256 maps - rgb, normals and specular. The system used for these timings (and used in the video) was a standard video-gaming desktop (Intel 7700k, 32Gb, and an NVidia 1070). The graphics card drove both the interactive previews and the GANs.

\subsection{Qualitative Comparisons}
\label{sec:qualitative_comparisons}
In this section, we qualitatively evaluate the style control of \systemName and compare our results to two end-to-end GANs: Pix2Pix~\cite{pix2pix} and BicycleGAN~\cite{zhu2017multimodal}. Similar quantiative evaluations and comparisons, based on two perceptual studies, are provided in the next section.

\begin{figure*}[t]
    \centering
    \includegraphics[width=\textwidth]{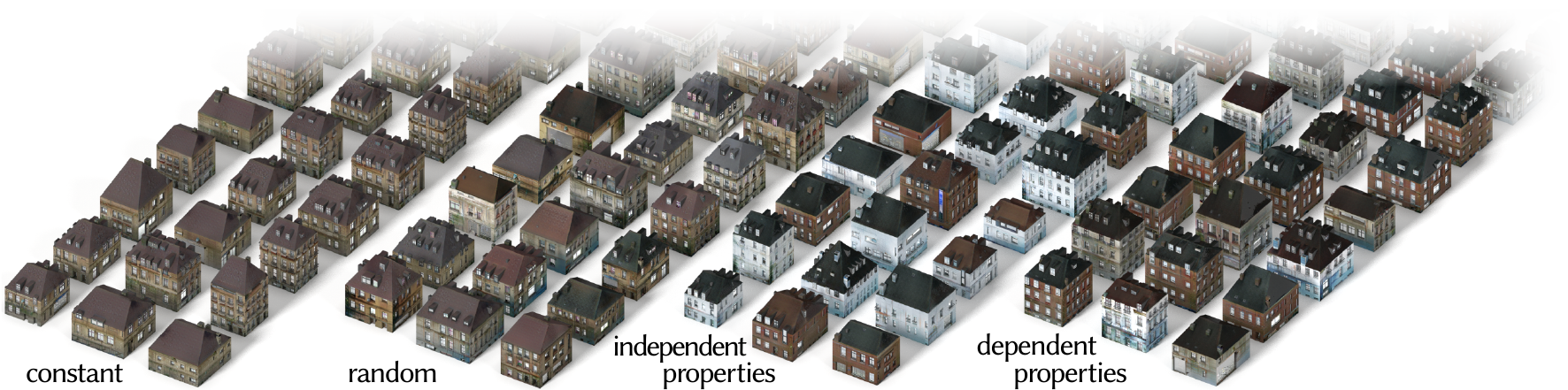}
    \caption{Different types of style distributions. From left to right, a constant style is used for all houses, a style chosen randomly from the style prior, a style sampled independently for all building properties, and a dependently sampled property style. Note that in the last two columns, there are several separate modes for properties, such as wall and building color, that are either mixed randomly for each building (third column) or sampled dependently (last column).}
    \label{fig:eval_style_control}
\end{figure*}

\begin{figure}[b]
    \centering
    \includegraphics[width=\columnwidth]{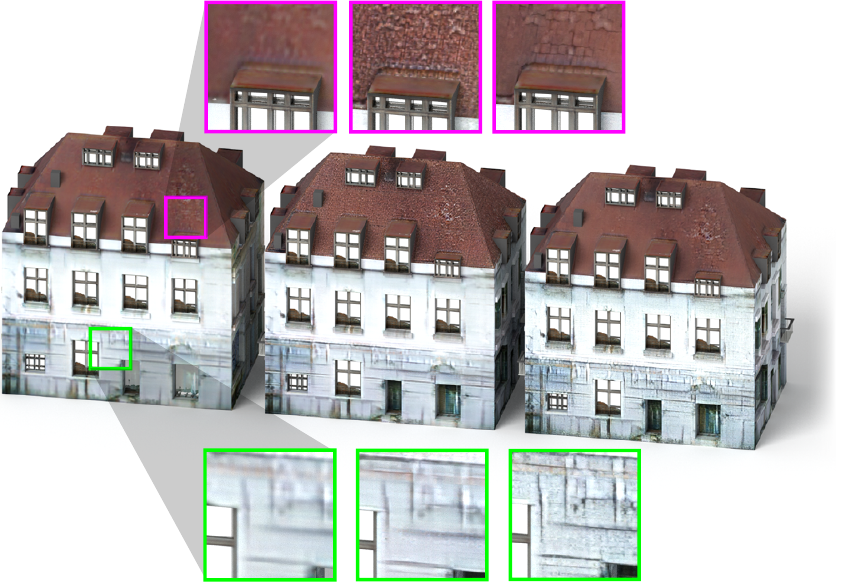}
    \caption{Different super-resolution styles. The original low-resolution facade and roof textures are shown on the left; the middle and right buildings show two different super-resolutions styles, resulting in different textures for the roof tiles or stone wall.}
    \label{fig:super_style}
\end{figure}

To evaluate the effects of style control on our outputs, we compare four different types of style distributions applied to similar mass models in Figure~\ref{fig:eval_style_control}. On the left, a constant style vector is used for all houses. This results in buildings with similar style, which may, for example, be required for row houses or blocks of apartment buildings. A fully random style is obtained by sampling the style prior, a standard normal distribution. The diversity in this case approximates the diversity in our training set. Using a Gaussian mixture model (GMM) as the style distribution, as described in Section~\ref{sec:style_control}, gives more control over the style. Note the two modes of white and red buildings in the third column, corresponding to two reference style images. The style for different building properties, such as roof and facade textures, is sampled independently, resulting in randomly re-mixed property styles, both white and red facades, for example, are mixed with black and red roofs. When specifying multiple sets of GMMs, one set is randomly picked for each house, allowing the user to model dependencies between property styles, as shown in the last column.

Examples of different super-resolution styles are shown in Figure~\ref{fig:super_style}. Since a low-resolution texture contains little information about fine texture detail, like the shingles on a roof or the stone texture on a wall, there is a diverse range of possible styles for this fine texture detail, as shown in the middle and right building.

\begin{figure*}[t]
    \centering
    \includegraphics[width=\textwidth]{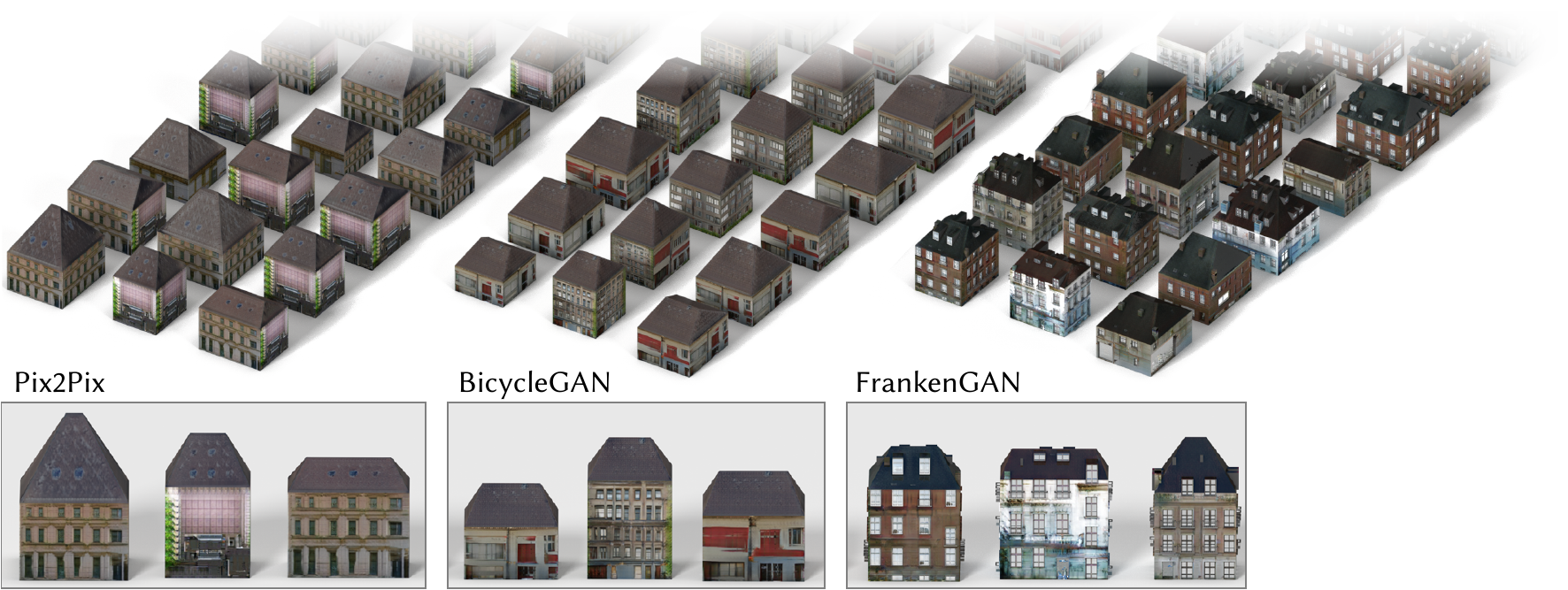}
    \caption{Qualitative comparison with end-to-end GANs. The left column shows results of Pix2Pix trained to transform empty facade and roof masks to textures. The middle column shows BicycleGAN trained similarly, while the last column shows our method. Note how Pix2Pix suffers from mode collapse, while BicyleGAN has less realistic window layouts and lacks scale and style consistency. \systemName provides better style control and our approach of splitting up the problem into multiple steps opens up several avenues to increase the realism of our models.}
    \label{fig:qual_comparison}
\end{figure*}

Figure~\ref{fig:qual_comparison} shows a qualitative comparison to Pix2Pix and BicycleGAN. We trained both end-to-end GANs on our facade and roof datasets, transforming empty facade and roof masks to textures in one step. As discussed in Section~\ref{sec:gan_architecture}, Pix2Pix, like most image-to-image GANs, has low output diversity for a given input image. In our case, the input image, being a blank facade or roof rectangle, does not provide much variation either, resulting in a strong mode collapse for Pix2Pix, shown by the repeated patterns such as the `glass front' pattern across multiple buildings. Like \systemName, BicycleGAN takes a style vector as input, which we set to a similar multi-modal distribution as in our method. There is more diversity than for Pix2Pix, but we observe inconsistent scale and style across different buildings, or across different facades of the same building, less realistic window layouts, and less diversity than in our results. Splitting the texture generation task into multiple steps allows us to provide more training signals, such as an explicit ground truth for the window layout, without requiring a very large network, regularize the intermediate results of the network and then use the intermediate results as label maps that can be used to generate geometric detail and assign materials. Additionally, giving the user more precise control over the style results in more consistent details across a building or parts of buildings. This allows us to generate more diverse and realistic building details. In the next section, we quantify the comparison discussed here with a perceptual study.

\subsection{\changed{Perceptual} Studies}

\begin{figure}[b]
    \centering
    \includegraphics[width=\columnwidth]{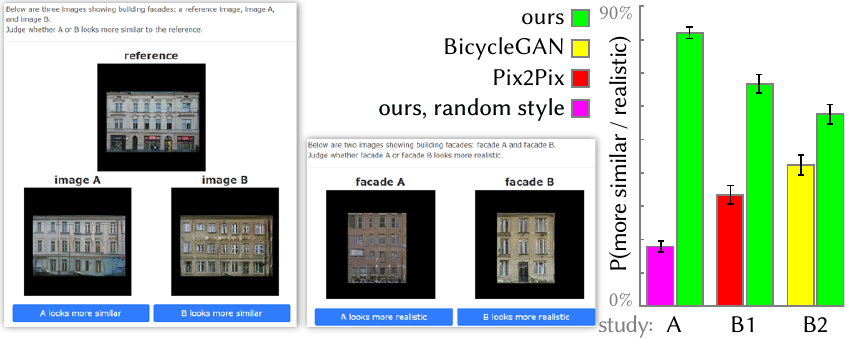}
    \caption{Perceptual studies comparing our method with/without style guidance (left) and to Pix2Pix~\cite{pix2pix} and BicycleGAN~\cite{zhu2017multimodal} (middle). The average probability for each method of being judged more similar to the reference or more realistic by the study participants is shown on the right. The black bars are $95\%$ confidence intervals.}
    \label{fig:study}
\end{figure}

We performed two perceptual studies to quantify the following questions:
\begin{enumerate}[label=(\Alph*)]
\item \emph{How visible are the effects of style and scale control?}
\item \emph{How does the realism of \systemName compare with that of Pix2Pix~\cite{pix2pix} and BicycleGAN~\cite{zhu2017multimodal}?}
\end{enumerate}
To investigate question A, we tested if participants could reliably tell which of two generated facades was guided by a given reference style facade. The style and scale was randomized for the other facade.
Question B was investigated by comparing facades generated by \systemName and one of the other two methods side-by-side and asking participants to compare the realism of the two facades.
We performed both studies in Amazon Mechanical Turk (AMT). Screenshots of both studies are shown in Figure~\ref{fig:study}; \changed{the assignment of images to the left and right positions was randomized.}

For study A, we created $172$ triples of facades, each consisting of a reference facade, a facade with style and scale given by the reference facade, and a third facade with randomized style. Participants were asked which of the two facades was more similar to the reference. Each triple was evaluated by an average of $11$ participants, and a total of $116$ unique participants took part in the study. Figure~\ref{fig:study} (green vs. blue) shows the average probability of a participant choosing either the style-guided facade or the facade with randomized style as more similar to the reference facade. Our results show that style-guided facade was consistently chosen, implying good visibility of style and scale control.

The second study was performed in two parts. In the first part, we compare our method to Pix2Pix, and in the second part to BicycleGAN. Users were shown one facade generated with \systemName and one facade with the other method, trained end-to-end to transform facade masks to facade textures. Each pair was evaluated by an average of $17.6$ and $17.5$ users, for Pix2Pix and BicycleGAN, respectively, and there were $86$ unique participants in the study. Results are shown in Figure~\ref{fig:study} (red vs. blue and yellow vs. blue). \systemName was chosen as more realistic in $66.6\%$ of the comparisons with Pix2Pix and $57.6\%$ of the comparisons for BicycleGAN. $95\%$ confidence intervals are shown as small bars in Figure~\ref{fig:study}. Note that this advantage in realism is in addition to the advantage of of having fine-grained style and scale control and obtaining label maps as intermediate results that are necessary to generate 3D details.

\subsection{Limitations}

\begin{figure}[b]
    \centering
    \includegraphics[width=\columnwidth]{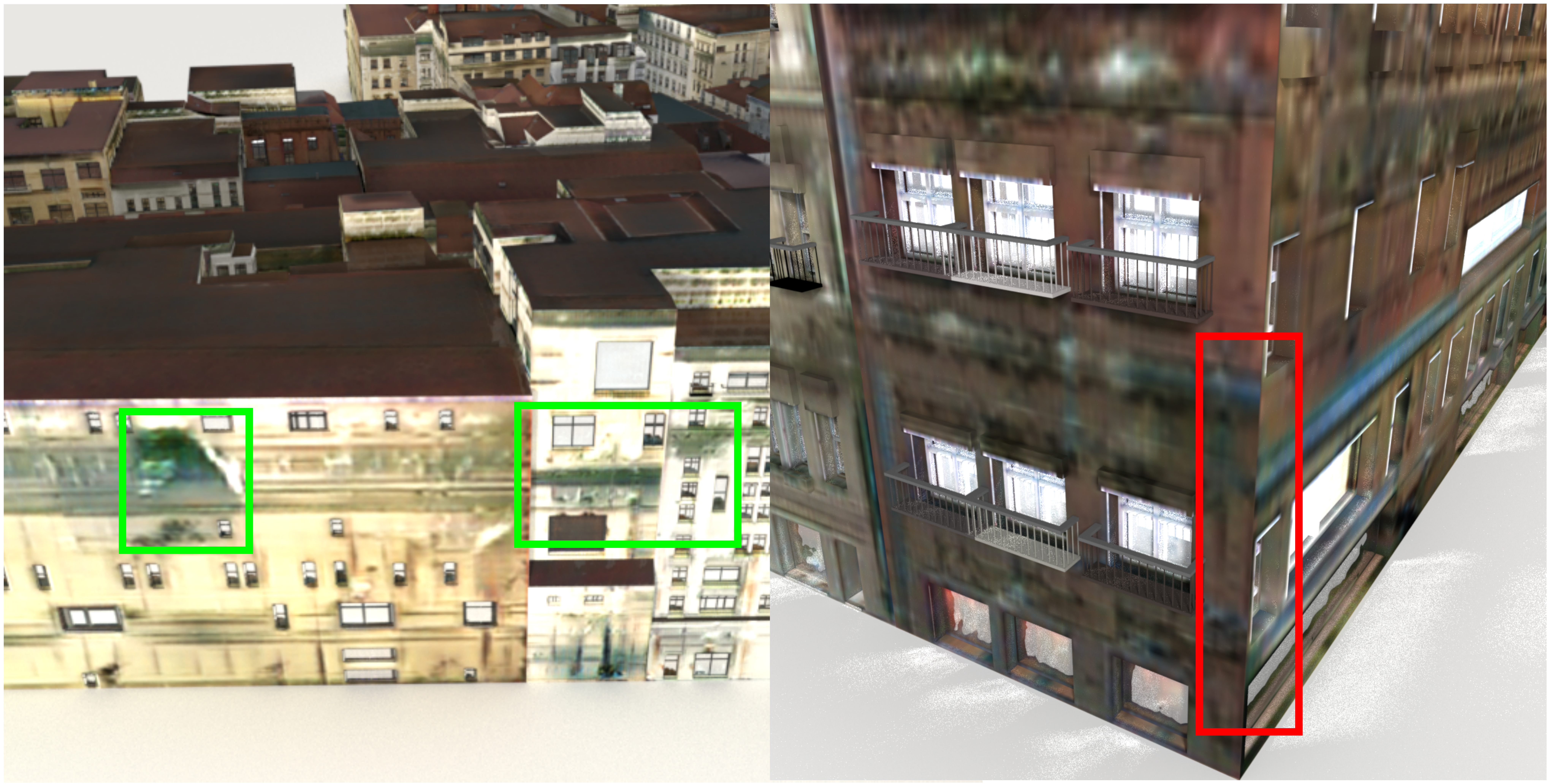}
    \caption{Limitations. Green: the system is prone to generating green vegetation over missing windows. Red: texture discontinuities at boundaries.}
    \label{fig:limitations}
\end{figure}

There are also some limitations to our framework. To replicate the framework from scratch requires extensive training data.
%while the GANs themselves is inherently difficult and requires experience.
%Exploring the training-parameter space of each dataset can involve several training runs.
Similar to other GANs, our results look very good from a certain range of distances, but there is a limit to how close it is possible to zoom in before noticing a lack of details \changed{or  blurriness}. This would require the synthesis of displacement maps and additional material layers in addition to textures. Data-driven texturing is inherently dependent on representative datasets. For example, our facade label dataset had missing windows due to occlusions by trees and other flora. We thus occasionally see missing windows in our results. We chose not to fill in windows in the regularisation stage.
%try to find these windows in the regularisation stage.
The facade-texturing network then learned to associate these missing windows with flora, and dutifully added green "ivy" to the building walls (Figure~\ref{fig:limitations}, left). Finally, our system uses a shared style to synchronize the appearance of adjacent building facades. However, this compact representation does not contain sufficient detail to guarantee seamless textures at boundaries (Figure~\ref{fig:limitations}, right).

%% file: sections/conclusion.tex
\section{Conclusion}

We presented \systemName, a system for adding realistic geometric and texture details on large-scale coarse building mass models guided by example images for style. Our system is based on a cascaded set of GANs that are individually controlled by generator vectors for style synchronization. We evaluated our system on a range of large-scale urban mass models and qualitatively evaluated the quality of the detailed models via a set of user studies.

The most direct way to improve the quality of output generated by \systemName is to retrain the GANs using richer and more accurately annotated facade and roof datasets and also to extend the system to generate street textures and furniture, as well as trees and botanical elements. This is a natural improvement as many research efforts are now focused on capturing high quality facade data with improved quality annotations. 
A more non-trivial extension is to explore a GAN-based approach that directly generates textured 3D geometry. 
The challenge here is to find a way to compare rendered procedural models with images based on photographs that also includes facade structure rather than just texture details.
Finally, it would be interesting to combine our proposed approach to author both realistic building mass models along with their detailing directly from rough input from the user in the form of sketches~\cite{Nishida:2016:ISU:2897824.2925951} or high-level functional specifications~\cite{Peng:2016:CND}.

\begin{acks}
This project was supported by an \grantsponsor{ERC}{ERC}{} Starting Grant (SmartGeometry \grantnum[]{ERC}{StG-2013-335373}), \grantsponsor{KAUST-UCL}{KAUST-UCL}{} Grant (\grantnum[]{KAUST-UCL}{OSR-2015-CCF-2533}), \grantsponsor{ERC}{ERC}{} PoC Grant (\grantnum[]{ERC}{SemanticCity}), the \grantsponsor{KOSR}{KAUST Office of Sponsored Research}{} (\grantnum[]{KOSR}{OSR-CRG2017-3426}), Open3D Project (\grantsponsor{EPSRC}{EPSRC}{} Grant \\ \grantnum[]{EPSRC}{EP/M013685/1}), and a \grantsponsor{Google}{Google}{} Faculty Award (\grantnum[]{Google}{UrbanPlan}).
\end{acks}